\documentclass{article}

\usepackage{graphicx}
\usepackage{float}

\def\bq{\begin{eqnarray}}
\def\eq{\end{eqnarray}}
\def\vf{\varphi}

\begin{document}

\begin{center}
{\bf {\LARGE Unruh effect without Rindler horizon}}

\end{center}

\begin{center}
Nistor Nicolaevici
\end{center}

\begin{center}
{\it Department of Physics,
The West University of Timi\c soara, V. P\^arvan 4,
300223, Timi\c soara, Romania}
\end{center}

\begin{abstract}

We investigate the Unruh effect for a massless scalar field in the two dimensional
Minkowski space in the presence of a uniformly accelerated perfect mirror, with
the trajectory of the mirror chosen in such a way that the mirror completely
masks the Rindler horizon from the space-time region of interest. We find that
the characteristic thermodynamical properties of the effect remain unchanged, i.e.
the response of a uniformly co-accelerated Unruh  detector and the distribution of
the Rindler particles retain their thermal form. However, since in this setup
there are no unobserved degrees of freedom of the field, the thermal statistics of
the Rindler particles is inconsistent with an initial pure vacuum, which
leads us to reconsider the problem for the more physical case when the mirror is
inertial in the past. In these conditions we find that the distribution of the Rindler
particles is non-thermal even in the limit of infinite acceleration times, but an
effective thermal statistics can be recovered provided that one restricts to the
expectation values of smeared operators associated to finite norm Rindler
states. We explain how the thermal statistics in our problem can be understood in
analogy with that in the conventional version of the effect.

\end{abstract}


\section*{1. Introduction}

One of the central results when considering quantum field theory in the general
relativistic framework is the Unruh effect, according to which a uniformly accelerated
observer in Minkowski space perceives the Minkowski  vacuum as a thermal bath at
temperature \cite{birr, cris}
\bq
T_U=\frac{\hbar a}{2\pi c k_B},
\label{unte}
\eq
where $a$ is the proper acceleration. The effect is usually presented in two versions. One
refers to the distribution of particles as defined in the co-accelerated (Rindler) frame,
which turns out to be identical with that in a thermal state of the field. The other
considers the response of a uniformly accelerated Unruh-deWitt particle detector, which
is also found to be of thermal form. The precise notion of thermality in this case is
expressed by the fact that the transition rate of the detector respects the detailed
balance or the KMS condition \cite{taka1}.

A natural question that can be asked about the Unruh effect, which led to various points of view
in the literature, is what explains the emergence of the thermodynamical properties. In discussing
this problem, it is important first to make clear what one means by ``Unruh effect.'' Restricting
to the two versions mentioned above,\, a useful distinction that was often made is as follows
(although not always under these names): since the concept of particle in quantum field theory
is a global construct, the first version can be called ``global''; the second version, which relies
on the local interaction between the detector and the field, can be called ``local.''

Another key concept in the discussion of the thermalization issue is that of the Rindler horizon,
i.e. the future event horizon of the uniformly accelerated observer. It appears that the existing
points of view can be largely distinguished according to the role assigned to the Rindler horizon.
Simplifying the picture, a fundamental question to be answered is whether the thermal properties
are intrinsically linked or not to the existence of the Rindler horizon. Without attempting a
history of the subject, the main views along with a number of significant results
can be summarized as follows.

In the global version of the effect, the horizon is an essential ingredient since it allows to
restrict the theory to the space-time region covered by the accelerated frame (usually a Rindler
wedge), and it introduces the important distinction between the observed and unobserved degrees
of freedom of the field. In this version the thermal properties are explained in terms of the
so-called thermalization theorem \cite{taka1, full, davi1}: the Bogolubov coefficients that mix
the Minkowski modes and the Rindler modes have a Boltzmann form, which for observations restricted
to a single Rindler wedge imply that the Minkowski vacuum is equivalent to a thermal state of the
field. The mixed state implied by the thermal statistics is then recognized to be a consequence
of ignoring the unobservable degrees of freedom of the field masked by the Rindler horizon, which
formally amounts to trace over these degrees of freedom. Thus, in this line of thought, thermality
and the Rindler horizon are intimately linked.

However, a different conclusion arises if one looks at the local version of the effect. An
illuminating\, gedanken-experiment at this point is as follows \cite{niko, schl, akhm, mart, rove}.
Consider a detector that accelerates only for a finite period of time, after which it follows
an inertial trajectory. It is easy to see that in these conditions the Rindler horizon does
not appear. Invoking causality and the locality of the field-detector interaction, one can
argue that for sufficiently large acceleration times the transition rates will become arbitrarily
close to the thermal rates for the infinitely accelerated trajectories.\footnote{One
can explicitly see this for a variety of trajectories in \cite{barb}.} This provides a simple
example which shows that, in the local version of the effect, thermality can also manifest
in the absence of the Rindler horizon (at least in an asymptotic sense).

It has to be mentioned that the view that the mechanism responsible for the thermal rates of the
detector  is intrinsically local and that thermality can be decoupled from the existence of the
Rindler horizon was firmly advocated more than two decades ago by Hu and collaborators in a series
of papers \cite{angl, rava1, rava2, koks}. Their conclusion was based on a calculation of the thermal
rates using the Feynman-Vernon influence functional method which relies entirely on local concepts and
thus emphasizes the local nature of the effect. The origin of thermality was explicitly identified in
the exponential scaling of the quantum vacuum fluctuations in the detector's proper frame, which
as\, a purely kinematic effect can also manifest in the absence of the Rindler
horizon.\footnote{The same underlying theory was shown to apply to related phenomena such as the
quantum radiation from accelerated mirrors and particle creation in curved space-times, with or
without horizons; see \cite{rava2, koks}.} In the author's own words, ``the exponential redshifting
is a more basic mechanism than the event horizon for thermal radiance'' \cite{rava2}. A somehow
similar explanation in local terms based on the Gaussian character of the vacuum
fluctuations was given earlier by Sciama et al. \cite{scia}, but without claiming the
independence of the effect from the acceleration horizon.

Evidence for the local nature of the effect is also provided by the recent works of Brenna et
al. \cite{bren} and Brown et al. \cite{brow}, who investigated the excitation of a detector
which accelerates inside a cavity, with different types of boundary conditions imposed on the
field at the\, edges of the cavity. They found that with a sufficiently long interaction time
and smooth switching of the interaction, the effective temperature defined by the reduced
density matrix of the detector shows a linear increase with its proper acceleration, a clear
sign of the Unruh effect.\footnote{For clarity, the effective temperature obtained in
\cite{bren, brow} does not reproduce the Unruh temperature (\ref{unte}). The distinction between
the two temperatures is made clear in \cite{lins}.} The notable fact is that the effective
temperature shows a negligible dependence on the type\, of boundary conditions, which
again indicates a decoupling of the effect from the global aspects of the problem, and
implicitly from the Rindler horizon.

Another example which illustrates the independence of the local effect from the Rindler
horizon was recently given by Rovelli and Smerlak \cite{rove}, who considered the response of
a detector which uniformly accelerates in the presence of a perfectly reflecting plane (mirror),
with the acceleration perpendicular to the plane. In this configuration it is possible to
choose the geometry so that no unobservable degrees of freedom of the field exist, which makes
the Rindler horizon irrelevant. Nevertheless, it turns out that for large acceleration times the
transition rates become arbitrarily close to the standard thermal rates in empty
space.\footnote{An identical conclusion was previously obtained in \cite{ohni, lang}, but without
discussing the connection with the Rindler horizon.} As stated in \cite{rove}, this shows that
``the thermal character of the transition rates of the uniformly accelerated detector cannot be
reduced to the effect of tracing out the modes behind the horizon,'' which essentially coincides
with the view in \cite{angl, rava1, rava2}

The aim of this paper is to add another item on the list in favour of the independence of the
Unruh effect from the Rindler horizon. The idea in our construction is to mask as in \cite{rove}
the Rindler horizon with a perfect mirror, but choosing this time the mirror to follow an
accelerated trajectory.\footnote{Beginning with the paper of Deutsch and Candelas \cite{deut}
there are a significant number of works which deal with quantum effects in the presence of
uniformly accelerated planes (see e.g. \cite{saha1, saha2}), but these papers have little in common
with our problem since they refer to the Rindler vacuum, whereas we consider the Minkowski
vacuum. For example, a co-accelerated detector remains unexcited in the Rindler vacuum,
which is not so in the Minkowski vacuum.} As we will see, a nice feature of this configuration
is that it allows to analyse both the local and global versions of the effect, much like in the
usual analysis in empty space \cite{birr}.

An aspect that somehow complicates the picture in \cite{rove} is that the transition rates depend
of time and that the thermal rates are exactly obtained only in the limit of infinite acceleration
times. It is easy to guess that the rates will be independent of time if the mirror follows a
co-accelerated trajectory with the detector, in the sense that the mirror accelerates such that
it appears for all times stationary in the detector's proper frame. We will examine the Unruh
effect in such a situation. For simplicity, we will focus on a massless scalar field in the two
dimensional Minkowski space. As is well known, co-accelerated trajectories with uniform accelerations
in the Minkowski plane can be identified with hyperbolae with the asymptotes defined by the edges
of the same Rindler wedge. This practically fixes the kinematics in our problem. We will work\,
as usual in the right Rindler wedge. A representation of the trajectories of the mirror and an
observer/detector in the configuration of interest is shown in Fig.~1. Note that for a perfect
mirror all the relevant degrees of freedom of the field are located on the right of the mirror.
It is clear that in these conditions no unobserved degrees of freedom exist and thus the Rindler
horizon is irrelevant.

Let us briefly state our main results. First, as expected, the transition rates of the detector
are constant along the trajectory and differ from the rates in empty space. The notable fact
is that they remain exactly thermal in the sense of the KMS condition,\footnote{We have reported
this result in a previous work \cite{nico}. The cited paper also contains a similar conclusion
for a detector which accelerates in the presence of contracting and expanding spherical mirrors
in four dimensions.} with the temperature defined by the standard temperature (\ref{unte}).
Second, the distribution of the Rindler particles is also exactly thermal. Actually, we will
find that the Bogolubov coefficients in the presence of the mirror are identical with the
coefficients in empty space (for the left moving modes).

\begin{figure}[H]

\centering

\includegraphics[width=0.9\linewidth]{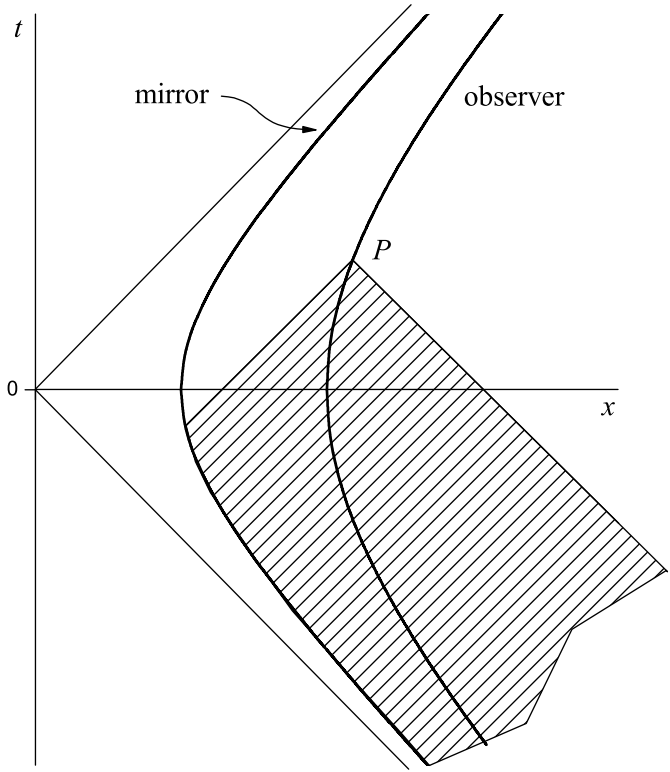}
\caption{The trajectories of a uniformly accelerated mirror and of a co-accelerated
observer on the right of the mirror. The dashed area represents the interior
of the past light cone of the point $P$ for a perfectly reflecting mirror.}

\end{figure}

However, the purely thermal statistics of the Rindler\, particles in the presence of the
mirror raises a problem. Since in our case there are no unobserved degrees of freedom of the
field, the mixed state implied by the thermal statistics in the Rindler frame is in
contradiction with a pure initial vacuum state. We ascribe this inconsistency to the unphysical
trajectories implied by the non-zero acceleration extending to infinite past times
$t\rightarrow-\infty$. One should also recall that a well-defined initial vacuum is
problematic in the presence of non-stationary boundaries, which is the case here.

Mathematically, the problem could be restated by saying that in the conditions of
interest the formal unitary transformation which exists between the Minkowski and the full
set of Rindler modes in empty space \cite{cris} breaks down. This is somehow unexpected
since here too the Minkowski and the Rindler modes are defined on a common region of space-time
in which they are both orthonormal and complete. We have no clear explanation for this fact.
The inconsistency might have to do with the calculation of the scalar products which define
the Bogolubov coefficients, in which we used integration by parts where we discarded as
usual the boundary terms at infinite distances. Perhaps these terms are the key to the
problem. In the last part of the paper we will discuss the thermalization of an initial
well-defined vacuum, in which we will see the special role  played by the degrees of
freedom at infinite distances, which seems to support this idea.

The inconsistency above invites to reconsider the problem for more physical trajectories
of the mirror. We will naturally choose trajectories with an acceleration that starts
at some finite time $t_0>-\infty$, before which the mirror moves inertially. We will repeat
the whole analysis for these trajectories. The picture in the new conditions can be summarized
as follows.

Not surprisingly, the transition rates of a detector which co-moves with the mirror (see text)
now become dependent on time, but\, for sufficiently large acceleration times they approach
the thermal rates in the previous calculation.  An analytical expression for the rates is
much more difficult to obtain in this case, so we will be content with numerical results. As
also to be expected, the Bogolubov coefficients for the new trajectories differ from the
coefficients in the first calculation, depending on the time $t_0$. Given the unproblematic
trajectories with the inertial mirror in the past, one can now expect that the Minkowski and
the Rindler modes are related via a unitary transformation, thus preserving the purity of the
initial vacuum in terms of Rindler states. Unfortunately, the complicated form of the coefficients
makes a direct verification of this fact an extremely difficult task.

A remarkable property of the new coefficients is that for $t_0\rightarrow -\infty$ they do not
reduce to the old ones, which can be seen to formally correspond to the case $t_0=-\infty$.
At first sight this seems counterintuitive, but the conclusion is welcomed: if the converse
were true, we would run again in contradiction with the purity of the initial vacuum state,
which has obviously nothing to do with the choice of the time $t_0$. However, from a more
physical point of view one can still expect that the thermal statistics defined by the old
coefficients can be somehow recovered when letting $t_0\rightarrow -\infty$. We will show that
this is indeed the case. The key point in establishing the connection is to restrict to the
expectation values of operators constructed from smeared creation and annihilation Rindler
operators associated to finite norm states, which is an otherwise perfectly natural restriction.
This will lead to a set of ``effective'' coefficients which in the limit will reproduce
the purely thermal coefficients in the first calculation.

Finally, having in mind that in our setup no unobserved degrees of freedom exist,
the emergence of the thermal statistics from the initial pure vacuum requires an explanation.
We will see that the restriction to smeared Rindler operators makes these operators insensitive
to the field at large Rindler distances from the mirror, and we will argue that the degrees of
freedom in this region are essential for maintaining the purity of the initial
vacuum. This will lead us to a close analogy with the usual picture for the Unruh effect
in empty space, in which the role of the unobserved Rindler wedge is played here by the
region at large Rindler distances from the mirror.

The paper is organized as follows. In the next section, we establish the form of the trajectories
and the quantum modes in the presence of the mirror. In Sec. 3 we obtain the transition rates and
the Bogolubov coefficients for the trajectories with
infinite acceleration times. In Sec. 4 we obtain the corresponding quantities for the
trajectories inertial in the past. We present our conclusions in Sec. 5. Technical matters
are relegated to Appendices A-C. In the rest of the paper we use natural units such that
$c=\hbar=k_B=1.$

\section*{2. Preliminaries}

Let us denote by $t, x$ the usual coordinates in the Minkowski plane. We choose the
uniformly accelerated trajectory of the mirror to be described by
\bq
\qquad
\qquad
x(t)=\left(t^2+1/a^2 \right)^{1/2},
\quad
t\in (-\infty, \infty),
\label{mirtra}
\eq
with $a>0$ the proper acceleration. It is useful to introduce the Rindler
coordinates $\eta,\, \xi$ defined by
\bq
t=\frac{1}{a}\, e^{a\xi} \sinh a\eta,
\quad
x=\frac{1}{a}\, e^{a\xi} \cosh a\eta.
\label{rincor}
\eq
We recall that\, these coordinates cover only the right Rindler wedge $x>\vert t\vert$.
The trajectory (\ref{mirtra}) in Rindler coordinates is simply
\bq
\xi=0.
\label{mtrrin}
\eq
while the co-accelerated trajectories with (\ref{mtrrin}) are
\bq
\xi=\mbox{constant},
\label{coactr}
\eq
with $\eta\in (-\infty, \infty)$. The region to the right of the mirror which is of interest here
corresponds to $\xi\in (0, \infty)$. The trajectories (\ref{coactr}) in
Minkowski coordinates as functions of the proper time $\tau$ read
\bq
t(\tau)=\frac{1}{a_\xi}\sinh a_{\xi}\tau,
\quad
x(\tau)=\frac{1}{a_\xi}\cosh a_\xi\tau,
\label{minmir}
\eq
where $a_\xi=a^{-a\xi}$ is the proper acceleration

The basic ingredient in our calculation are the quantum modes of the field.
We first recall the form of the modes in empty space. The positive frequency
Minkowski modes are (for both sets $k\in(-\infty, \infty)$)
\bq
\qquad
\varphi^M_\omega(t,x)=\frac{1}{\sqrt{4\pi\omega_k}}\, e^{ikx-i\omega t},
\quad \omega=\vert k\vert.
\qquad
\label{minmo0}
\eq
The Rindler modes are
\bq
\varphi^R_\omega(\eta,\xi)=\frac{1}{\sqrt{4\pi\omega_k}}\, e^{ik\xi-i\omega \eta},
\quad \omega=\vert k\vert.
\label{rinmo0}
\eq
Note that the Rindler modes are defined only within the right Rindler wedge. Both sets of
modes together with their complex conjugates are orthonormal and complete in their definition
domain, which allows to construct the quantum theory with the standard procedure \cite{birr}.

We now discuss the modes in the presence of the mirror. We ensure the perfect reflectivity
condition by imposing Dirichlet boundary conditions $\varphi =0$, which must hold everywhere
along the trajectory of the mirror. The Minkowski modes can be readily obtained using the
moving mirror model of Fulling and Davies \cite{birr, davi2}. We recall that
there are two sets of modes of special physical interest, i.e. the $in$ and the $out$ modes.
From physical reasons it is clear that the Minkowski modes to be used in our calculation are
the $in$ modes, i.e. the modes which describe an unperturbed wave in the infinite past.
Let us introduce the null coordinates
\bq
u=t-x,
\quad
v=t+x.
\label{nulcor}
\eq
An arbitrary trajectory of the mirror can be specified via the dependence
\bq
v=p\,(u).
\label{funcef}
\eq
The positive frequency Minkowski $in$ modes for an arbitrary motion
of the mirror in the region to the right of the mirror are \cite{birr}
\bq
\varphi^M_\omega(u,v)=\frac{1}{\sqrt{4\pi\omega}}
\left (e^{-i\omega v} - e^{-i\omega p\,(u)}\right),
\quad
\omega>0.
\label{minmod}
\eq
For the uniformly accelerated trajectories (\ref{mirtra}) the function (\ref{funcef})
is\footnote{We will systematically attach a bar to the quantities defined by these
trajectories.}
\bq
\bar p\,(u)=-\frac{1}{a^2 u}, \quad u\in (-\infty, 0).
\label{trajuv}
\eq
We will need (\ref{minmod}) in terms of the coordinates $\eta$, $\xi$. Using
\bq
u=-\frac{1}{a}\, e^{a (\xi-\eta)},
\quad
v=\frac{1}{a}\, e^{a (\xi+\eta)},
\label{nulrin}
\eq
one obtains
\bq
\bar \varphi^M_\omega(\eta, \xi)=
\frac{1}{\sqrt{4\pi\omega}}
\left(
e^{-i (\omega/a)\,e^{a(\xi+\eta)}}
-
e^{-i (\omega/a)\,e^{-a(\xi-\eta)}}
\right).
\label{expmin}
\eq

The Rindler modes in the presence of the mirror can be easily found since in
Rindler coordinates the mirror appears to be stationary at $\xi=0$. The positive
frequency solutions with the desired boundary conditions
are
\bq
\qquad
\bar \varphi^R_\omega(\eta, \xi)
=\frac{1}{\sqrt{4\pi\omega}}
\left (e^{-i\omega (\xi+\eta)} - e^{i\omega(\xi-\eta)}\right),
\quad
\omega>0.
\label{rinmod}
\eq
Both sets of modes (\ref{minmod}) and (\ref{rinmod}) together with their complex conjugates
are orthonormal and complete in the region to the right of the mirror.

We stress that the modes (\ref{expmin}) and (\ref{rinmod}) with $\eta \in (-\infty, \infty)$,
$\xi\in (0,\, \infty)$ are valid only for the trajectories (\ref{mirtra}) where it is essential
for $t$ to vary over the entire real axis. In Sec. 4 we will deal with accelerations
which begin only at a finite time $t_0$. The implications for the form of the modes are as
follows. For the Minkowski modes, the new trajectories will imply a different function $p\,(u)$
and thus a different expression (\ref{minmod}) (we will write the modes explicitly
in the due place). For the Rindler modes, we can arrange things so that they remain unchanged.
The argument is as follows. Note that in the right Rindler wedge the condition $t\geq 0$ is
equivalent to $\eta\geq 0$ and that this inequality selects the superior ``triangular'' sector
of the wedge. The essential observation is that, if the mirror starts to accelerate at a
negative time $t_0<0$, the picture in the space-time sector $\eta\geq 0$ is identical with
that for the trajectories with infinite acceleration times (\ref{mirtra}). We will always
assume that this is the case and concentrate on this space-time region. It
is then easy to see that the same modes (\ref{rinmod}) will appropriately describe the
particles in the presence of the mirror in the accelerated frame.

\section*{3. Accelerations beginning at $t\rightarrow -\infty$}

We consider in this section the case when the mirror and the detector uniformly
accelerate for all times $t\in (-\infty, \infty)$. We are interested in (1) the
transition rates of the co-accelerated detector, and (2) the distribution of the
Rindler particles in the presence of the mirror. We will use the standard
Unruh-DeWitt detector model \cite{birr}.  The co-accelerated trajectory of the
detector $(D)$ will be fixed by its spatial Rindler coordinate $\xi_D>0$. A
representation of the trajectories of the mirror and of the detector is shown
in Fig.~2.

\begin{figure}[H]

\centering

\includegraphics[width=0.9\linewidth]{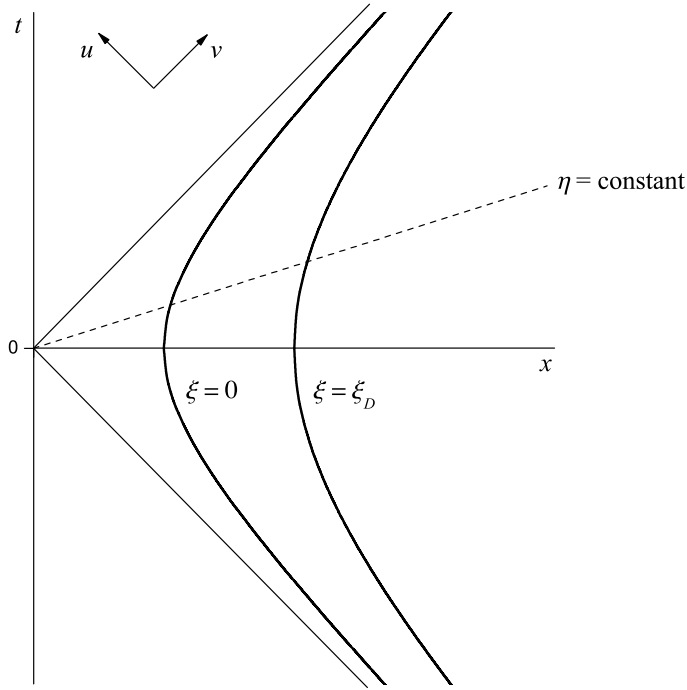}

\caption{The trajectories of the mirror and of a co-accelerated
detector as lines $\xi=\,$constant in Rindler coordinates.}

\end{figure}

{\large {\bf 3.1 The response of the particle detector}}
\\
\\
Let us denote by ${\bf x}(\tau)$ the trajectory of the detector in the Minkowski space, with $\tau$
the proper time. We recall that, in the special case when the field-detector system is invariant to
translations with respect to $\tau$, the transition rates from the ground state $E_0=0$ to an energy
level $E$ are \cite{birr}
\bq
\qquad {\cal R}(E)=\int_{-\infty}^{\infty}
d \Delta\tau\, e^{-i E\Delta \tau}
D^+({\bf x} (\tau_1), {\bf x}(\tau_2)),
\quad
\Delta \tau\equiv\tau_1-\tau_2,
\label{detrat}
\eq
where $D^+({\bf x}_1, {\bf x}_2)$ is the positive frequency Wightman function
in the state of the field. (We neglect as usual a factor that only depends on the structure of
the detector.) The translational invariance of the system is expressed by the condition
\bq
D^+({\bf x}(\tau_1), {\bf x}(\tau_2))=D^+(\Delta \tau).
\label{invtra}
\eq
It will become soon clear that (\ref{invtra}) is respected in the case of interest
(see eqs. (\ref{d0})-(\ref{d2}) below).

As already mentioned, the role of the usual Minkowski vacuum in the empty space calculation\, is now
played by the vacuum defined by the $in$ modes (\ref{minmod}). The positive frequency Wightman function
determined by these modes for an arbitrary trajectory of the mirror is \cite{birr}
\bq
\quad D^+(u,v;\, u^\prime, v^\prime)
=-\frac{1}{4\pi}
\ln
\frac{(p\,(u)-p\,(u^\prime)-i\varepsilon) (v-v^\prime-i\varepsilon)}
{(v-p\,(u^\prime)-i\varepsilon)(p\,(u)-v^\prime-i\varepsilon)},
\quad  \varepsilon\rightarrow 0_+.
\label{wiggen}
\eq
For the trajectories (\ref{trajuv}) this translates into\footnote{We absorbed as commonly done
a positive factor into the $i\varepsilon$ term. It is essential at this step to use the fact
that everywhere in the right Rindler wedge $u<0$.}
\bq
D^+(u,v;\, u^\prime, v^\prime)
=-\frac{1}{4\pi}
\ln (u-u^\prime-i\varepsilon) (v-v^\prime-i\varepsilon)\,\,
\qquad\qquad
\nonumber
\\
-\frac{1}{4\pi}
\ln \frac{-1/a^2}{(uv^\prime+1/a^2-i\varepsilon)(v u^\prime+1/a^2+i\varepsilon)}.
\label{wigacc}
\eq
Note that in the first logarithm we isolated the positive frequency Wightman function
for the free field in empty space. We have to evaluate (\ref{wigacc}) along the trajectory
of the detector, which is given by (\ref{minmir}) where we have to set $\xi=\xi_D$. Introducing
the proper acceleration
\bq
a_D=a e^{-a\xi_D},
\label{detacc}
\eq
the detector's trajectory in null coordinates is
\bq
u(\tau)=-\frac{1}{a_D}\, e^{-a_D \tau},
\quad
v(\tau)=\frac{1}{a_D}\, e^{a_D \tau}.
\label{tranul}
\eq
Inserting (\ref{tranul}) into (\ref{wigacc}) one finds
\bq
D^+({\bf x}(\tau_1), {\bf x}(\tau_2))= D^+_0(\Delta \tau) + D^+_1(\Delta \tau),
\label{d0}
\eq
where (the terms $D_0^+$ and $D_1^+$ correspond to the two logarithms in (\ref{wigacc}))
\bq
 D_0^+(\Delta\tau)
 =-\frac{1}{4\pi}
\ln
\left\{
\sinh^2 \left(\frac{a_D\Delta \tau}{2}-i\epsilon\right)
\right\},
\label{d1}
\eq
\bq
D_1^+(\Delta\tau)
=\frac{1}{4\pi}
\ln
\left\{
\sinh \left(\frac{a_D\Delta \tau}{2}
+a\xi_D -i\epsilon\right)
\sinh \left(\frac{a_D\Delta \tau}{2}
-a\xi_D-i\epsilon\right)
\right\}
\label{d2}.
\eq

A convenient way to evaluate (\ref{detrat}) with the integrand defined by (\ref{d0})-(\ref{d2})
is as follows \cite{taka1}. The first step is to eliminate the logarithms with an integration by
parts (this can be justified with an adiabatic decoupling of the interaction at $\tau\rightarrow
\pm\infty$), which replaces the functions $\ln \sinh z$ by coth$z$. The
integrand can then be rewritten with
\bq
\coth (z - a) =\sum_{n=-\infty}^{\infty}\frac{1}{z - a +in\pi},
\eq
which allows to use a contour integration in the complex $\Delta \tau$ plane. The integrals are
essentially the same as in the standard calculation for the uniformly accelerated detector in
empty space (see \cite{nico} for details). We recall that in these conditions the rates
are ($\Theta$ is the unit step function)
\bq
\hat {\cal R}(E)=\frac{1}{\vert E\vert}
\,
\left(
\Theta(E)\,\frac{1}{e^{\beta_D E}-1}+\Theta(-E)\,\frac{e^{\beta_D E}}{e^{\beta_D E}-1}
\right),
\quad
\beta_D =\frac{2\pi}{a_D}.
\label{thra}
\eq
Expressing our rates
in terms of
(\ref{thra}) the result is \cite{nico}
\bq
\bar {\cal R}(E)=\chi(E,\, \xi_D)\, \hat {\cal R}(E),
\label{tramir}
\eq
where
\bq
\chi(E, \chi_D)=1-\cos \left (2E\xi_D\, e^{a\xi_D}\right).
\label{frofac}
\eq

Let us make a few comments on (\ref{tramir}). The unit term in (\ref{frofac}) corresponds
to the standard rates (\ref{thra}), while the second term contains the perturbation due to
the mirror. Notice the oscillatory behaviour of this term with the detector's coordinate
$\xi_D$, which can be naturally seen as an interference effect between the incident
and the reflected vacuum fluctuations in the presence of the mirror. (The oscillatory
behaviour of the rates with respect to the detector's position appears to be a constant
feature in the presence of boundaries, see \cite{lang, davi3, ford}.) One may also observe
that the perturbation term does not vanish for infinite mirror-detector distances
$\xi_D\rightarrow \infty$, which can be traced to the one-dimensional nature of
the problem. (The perturbation terms in the four-dimensional examples in \cite{nico}
vanish in the same limit.)

It is\, now easy to check that the perturbed\, rates remain thermal. The thermal property
in terms of the KMS condition is expressed by \cite{taka1}
\bq
{\cal R}(E)/{\cal R}(-E) = e^{\beta E},
\label{kmscon}
\eq
with $\beta$ the inverse temperature. Since (\ref{kmscon}) is respected by the thermal rates
(\ref{thra}) with $\beta=\beta_D$, it is immediate from (\ref{tramir}) that the perturbed
rates will respect the same condition provided that
\bq
\chi(E, \xi_D)=\chi(-E, \xi_D).
\label{relchi}
\eq
One sees from (\ref{frofac}) that relation (\ref{relchi})\, indeed holds. A more direct
way to arrive to the same\, conclusion is to recall that the KMS condition is equivalent
to \cite{taka1} (with $\Delta \tau$ seen as a complex variable)
\bq
D^+(\Delta \tau)=D^+(-\Delta \tau -i\beta),
\eq
which is separately respected by  ({\ref{d1}) and (\ref{d2}).
\\
\\
{\large {\bf 3.2 The distribution of Rindler particles}}
\\
\\
We now consider the distribution of the Rindler particles in the $in$ vacuum in the
presence of the mirror. The answer is contained in the Bogolubov coefficients that
mix the Rindler and the Minkowski $in$ modes. We will focus on the beta coefficients,
the calculation for the alpha coefficients being essentially the same. The beta
coefficients are given by \cite{cris}
\bq
\beta
(I, i)=(\varphi_I^{R\,*},\, \varphi_{i}^M),
\label{betcoe}
\eq
where the parentheses denotes the usual scalar product for scalar fields. We first
recall the result in empty space. In this case the scalar product in Rindler coordinates
is
\bq
(\varphi_1, \varphi_2)=i\int_{-\infty}^\infty d\xi\,
\varphi_1^* \stackrel{\leftrightarrow}{\partial}_\eta \varphi_2,
\label{scapro}
\eq
where the integration extends over an arbitrary hypersurface $\eta =\,$constant.
Using the modes in Sec. 2 and choosing $\eta=0$ one finds with an integration by
parts that for the left-moving modes $k, k^\prime <0$ the coefficients are \cite{cris}
\bq
\beta(\omega, \omega^\prime)=-\frac{1}{2\pi} \sqrt {\frac{\omega}{\omega^\prime}}
\int_{-\infty}^\infty d\xi\,
e^{-i\omega \xi -i (\omega^\prime/ a)\, e^{a\xi}}
\qquad\qquad\,\,
\nonumber
\\
\nonumber
\\
=-\frac{1}{2\pi a} \sqrt {\frac{\omega}{\omega^\prime}}
\left (\frac{a}{\omega^\prime}\right)^{-i\omega/a}
e^{-\pi \omega/2a}\, \Gamma (-i \omega/a),
\,\,\,\,
\label{frecoe}
\eq
where $\Gamma(z)$ is the Euler Gamma function. The analogous result for the
right-moving modes is given by the complex conjugate of (\ref{frecoe}). The
coefficients that mix the left- and right-moving modes are identically null.

We now consider the problem in the presence of the mirror. The field in this
case is defined only for $\xi\geq 0$ and the scalar product has to be rewritten
as
\bq
(\bar \varphi_1, \bar\varphi_2)=i\int_{0}^\infty d\xi\,
 \bar \vf_1^* \stackrel{\leftrightarrow}{\partial}_\eta \bar \vf_2,
\label{promir}
\eq
where the integration can be similarly taken on an arbitrary hypersurface $\eta =\,$constant
(the vanishing of the field at $\xi=0$ ensures that (\ref{promir}) remains independent
of $\eta$). A convenient way to perform the calculation is as follows. We observe that
the modes in Sec. 2 can be organized as
\bq
\bar \varphi^M_\omega(\eta, \xi)&=&F^{inc}_\omega(\xi+\eta)-F^{ref}_\omega (\xi-\eta),
\label{fg1}
\\
\bar\varphi^R_\omega(\eta, \xi)&=&G^{inc}_\omega(\xi+\eta)-G^{ref}_\omega (\xi-\eta),
\label{fg2}
\eq
where the form of the functions $F_\omega$ and $G_\omega$ can be read from (\ref{expmin})
and  (\ref{rinmod}). We further note that (the superscripts $inc$ and $ref$ are suppressed
with no loss of clarity)
\bq
\qquad
\partial_\eta F_\omega(\xi \pm\eta)\Big \vert_{\,\eta=0}
&=&
\pm F_\omega ^{\,\prime}(\xi),
\label{deri1}
\\
\quad
\partial_\eta G_\omega(\xi\pm \eta)
\Big \vert_{\,\eta=0}
&=&
-i\omega\, G_\omega(\xi),
\label{deri2}
\eq
where the prime denotes derivation with respect to the argument. We choose as before
the integration hypersurface $\eta=0$ in (\ref{promir}). The idea is to use integrations
by parts in order to eliminate the derivatives $F_\omega^\prime$ introduced by
(\ref{deri1}). The details of the calculation are given in Appendix A. The result in
terms of $F_\omega$ and $G_\omega$ is
\bq
\bar \beta(\omega, \omega^\prime)=
(-2\omega)\times \int_0^\infty d\xi
\left\{
G_\omega^{inc}(\xi)\, F_{\omega^\prime}^{inc}(\xi)+
G_\omega^{ref}(\xi)\, F_{\omega^\prime}^{ref}(\xi)
\right\}.
\label{genint}
\eq
Note that (\ref{genint}) contains no mixing between the incident and the reflected
components of the modes. Using the explicit form of $F_\omega$ and $G_\omega$ the
integral translates into
\bq
\bar \beta(\omega, \omega^\prime)=
-\frac{1}{2\pi} \sqrt {\frac{\omega}{\omega^\prime}}
\int_0^\infty d\xi\,
\left
\{
e^{-i\omega \xi -i (\omega^\prime/ a)\, e^{a\xi}}+
e^{i\omega \xi -i (\omega^\prime/ a)\, e^{-a\xi}}
\right \}.
\eq
We now change in the second term in the bracket the sign of the integration
variable $\xi\rightarrow -\xi$ and observe that the two terms can be summed into
a single integral, which leads to
\bq
\bar \beta(\omega, \omega^\prime)=
-\frac{1}{2\pi} \sqrt {\frac{\omega}{\omega^\prime}}
\left \{
\,\int_{0}^\infty \dots
+
\,\int_{-\infty}^0 \dots
\right \}\,\,\,\,
\nonumber
\\
\nonumber
\\
=-\frac{1}{2\pi} \sqrt {\frac{\omega}{\omega^\prime}}
\int_{-\infty}^\infty d\xi\,
e^{-i\omega \xi -i (\omega^\prime/ a)\, e^{a\xi}}.
\label{mircoe}
\eq
Comparing with (\ref{frecoe}) one sees that the two integrals with respect to $\xi$
are identical. Thus the  conclusion is that the two set of coefficients coincide,
\bq
\bar \beta(\omega, \omega^\prime)=\beta(\omega, \omega^\prime).
\label{idenb}
\eq

The same conclusion can be obtained for the alpha coefficients \cite{cris}
\bq
\alpha (I, i) =(\varphi_I^{R}, \, \varphi_{i}^M)^*.
\label{alptcoe}
\eq
There is actually no need to repeat the calculation if one observes from (\ref{betcoe})
and (\ref{alptcoe}) that for the modes in the case of interest (this includes
the modes in Sec.~4) the alpha and beta coefficients are related via\footnote{With the
extra rule that in the factors $\sqrt\omega$\, in front of the integrals the sign of
$\omega$ remains unchanged.}
\bq
\alpha(\omega,\omega^\prime)=-\beta(-\omega,\omega^\prime)^*.
\label{alfbet}
\eq
It is then immediate from (\ref{idenb}) that (using a similar notation)
\bq
\bar \alpha(\omega, \omega^\prime)=\alpha(\omega, \omega^\prime).
\label{idena}
\eq
Note that thanks to (\ref{alfbet}) all expressions for the beta coefficients we will obtain
here can be readily translated for the alpha coefficients.

As a side mention, it is interesting to remark from (\ref{genint})-(\ref{mircoe}) that the
identity between the two sets of coefficients can be seen to be a consequence of the following
facts: $(i)$ the contribution of the incident components of the modes in
$\bar\beta(\omega,\omega^\prime)$ (i.e. the term $inc$-$inc$ in (\ref{genint}) or the first integral
in (\ref{mircoe})) is identical with the contribution of the region $\xi>0$ in
$\beta(\omega,\omega^\prime)$, $(ii)$ the same is true for the contribution of the reflected
components (i.e. the term $ref$-$ref$ in (\ref{genint}) or the second integral in
(\ref{mircoe})) and that of the region $\xi<0$ in the same coefficients, and $(iii)$ there is no
mixing between the incident and reflected components of the modes, similar to the non-mixing
between the left- and right-moving modes in empty space. Is it as if in the presence of the
mirror the ``missing part'' of the modes in the region on the left of the mirror $\xi<0$ is
compensated by  their component reflected back in the region $\xi>0$.

We now recall the well-known fact \cite{birr, cris} that for the free field in the empty Minkowski
space the form of the Bogolubov coefficients imply that the Minkowski vacuum appears to the Rindler
observers as a thermal state with temperature $1/\beta=a/2\pi$. The identities (\ref{idenb}) and
(\ref{idena}) show that the same is valid in the presence of the mirror. In particular, the
distribution of the Rindler particles in the new conditions remains purely thermal.

\section*{4. Accelerations beginning at $t_0>-\infty$}

As we have remarked in Sec. 1, the identity between the two sets of coefficients
raises a problem. Since in the presence of the mirror there are no unobserved degrees
of freedom of the field, the initial Minkowski vacuum should remain pure when expressed
in terms of Rindler states. However, this is in contradiction with the exact thermality
implied by the coefficients in empty space. A simple way out of the problem is to identify
the source of the inconsistency in the unphysical trajectories with a non-zero
acceleration which extends to $t\rightarrow -\infty$. We will thus reconsider our\,
analysis assuming that the mirror is inertial before some finite time $t_0$ in the past,
with $t_0<0$ as discussed in Sec. 2.

As a first step we establish the trajectories of the mirror. We assume that for $t>t_0$ the
evolution is according to the same accelerated trajectories (\ref{mirtra}). We denote by $J$
the junction point at the time $t=t_0$. We naturally choose the trajectory
to be of class ${\cal C}^1$ at this point. Let us denote the new trajectory function (\ref{funcef})
by\footnote{We will attach a tilde to the quantities defined by the new trajectories of the mirror.}
\bq
v=\tilde p\,(u).
\label{trabar}
\eq
A simple calculation shows that
\bq
\tilde p\,(u)=
\left
\{
\begin{array}{ll}
\tilde p_-(u)\quad &\mbox{if}\quad u < u_0
\\
\tilde p_+(u)\quad &\mbox{if}\quad  u \geq u_0,
\end{array}
\right.
\label{trainp}
\eq
with
\bq
\tilde p_-(u)=\frac{u-2u_0}{a^2 u_0^2},
\quad
\quad
\tilde p_+(u)=-\frac{1}{a^2 u},
\eq
where $u_0<0$ is the retarded null coordinate of the point $J$. The connection
with the time $t_0$ is
\bq
t_0=\frac{u_0}{2}-\frac{1}{2a^2 u_0}.
\label{contiz}
\eq
A diagram which illustrates the trajectory of the mirror in the new conditions is shown
in  Fig.~3 (the details on the diagram will be helpful at a later time). As discussed
in Sec. 2 the new Minkowski $in$ modes
are given by
\bq
\tilde \varphi^M_\omega(u,v)=\frac{1}{\sqrt{4\pi\omega}}
\left (e^{-i\omega v} - e^{-i\omega \tilde p\,(u)}\right),
\label{minine}
\eq
while the Rindler modes remain unchanged,
\bq
\tilde \varphi^R_\omega(u,v)=\bar \varphi^R_\omega(u,v).
\label{rinine}
\eq

{\large {\bf 4.1. The response of the comoving detector}}
\\
\\
In order to stay as close as possible to the calculation in Sec. 3.1, we will assume that
the detector follows a comoving trajectory with the mirror. We mean by this that the
trajectory is such that the detector appears for $all$ times stationary in the proper
frame of the mirror, including during the stage of inertial motion. Let us
consider the hypersurface $\eta=\eta_0$ that passes through the point $J$ (see Fig.~3).
It is then\, immediate from the diagram that our condition implies that (1) above
this line the trajectory of the detector has to coincide with the co-accelerated trajectory
in Sec. 3.1, while (2) below the line the trajectory has to be inertial,
with the velocity equal to the velocity of the mirror for $t<t_0$.

\begin{figure}[H]

\centering

\includegraphics[width=0.9\linewidth]{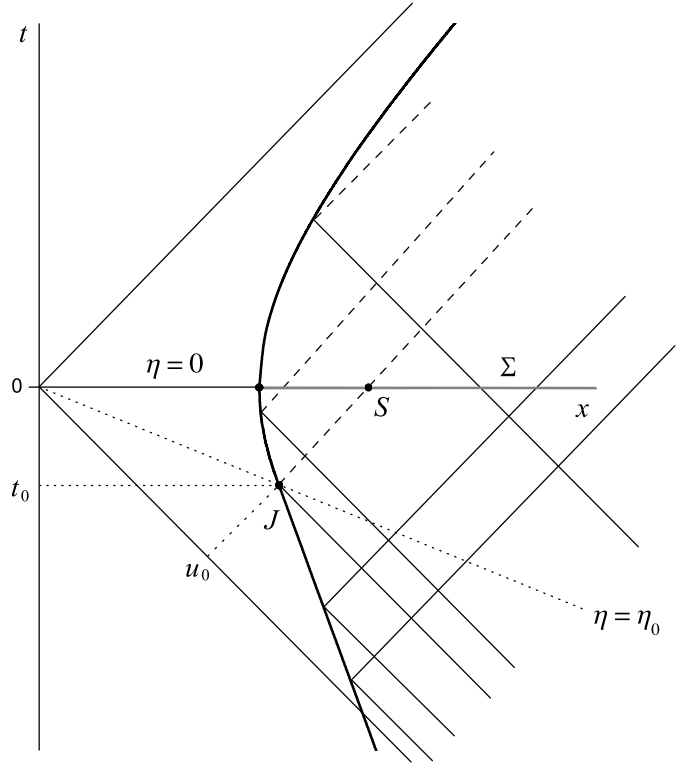}

\caption{Trajectory of the mirror with constant velocity for $t<t_0$. The lines at
$45^\circ$ represent light rays. The dashed lines represent rays reflected by the
mirror on the accelerated part of the trajectory.}

\end{figure}

The picture above can be slightly simplified as follows. Let us denote by $\tau_0$ the proper
time of the detector at which its trajectory intersects the hypersurface $\eta=\eta_0$.
Adopting the point of view of an observer attached to the detector, it is clear that in
the detector's proper frame the mirror appears
to be stationary for $\tau<\tau_0$. Note further that by the relativistic invariance of the
theory, the initial vacuum in the presence of the inertial mirror coincides with the vacuum
state in the observer's proper frame. It is then easy to see that for $\tau<\tau_0$ the
detector will behave like a stationary detector in the Minkowski vacuum in the presence of a
stationary mirror (thus having a zero excitation rate), while for $\tau> \tau_0$ the transition
rates will depend on $\tau$ only via the difference $\tau-\tau_0$. This means that, without
loss of generality, we can choose $\tau_0=0$. This is equivalent with $t_0=0$, in which case
both the mirror and the detector are fixed for all negative times $t<0$. In this section we will
focus on these trajectories.

In the conditions above, the trajectory function (\ref{trainp}) simplifies to
\bq
\tilde p\,(u)=
\left
\{
\begin{array}{ll}
\,\,u+2/a
\quad &\mbox{if}\quad u< -1/a
\\
-1/(a^2 u)
\quad &\mbox{if}\quad u \geq  -1/a.
\end{array}
\right.
\label{tmp}
\eq
The trajectory of the detector as a function of $t$ is
 \bq
 \ x(t)=
 \left
 \{
 \begin{array}{ll}
 \,\,1/a_D
 \quad &\mbox{if}\quad t< 0
 \\
\left(t^{\,2}+1/a_D^{2}\right)^{1/2}
\quad &\mbox{if}\quad t \geq 0,
 \end{array}
 \right.
\eq
or as a function of $\tau$ in null coordinates ($a_D$ is defined in (\ref{detacc}))
\bq
\ u(\tau)=
\left
\{
\begin{array}{ll}
\tau-a_D^{-1}
\\
-a_D^{-1}\, e^{-a_D \tau},
\end{array}
\right.
\quad
v(\tau)=
\left
\{
\begin{array}{ll}
\tau+a_D^{-1}
\quad &\mbox{if}\quad \tau< 0
\\
a_D^{-1}\, e^{a_D \tau}
\quad &\mbox{if}\quad \tau \geq  0.
\end{array}
\right.
\label{tdp}
\eq

It is clear that the transition rates for the new trajectories will be time-dependent
quantities. This means that we have to consider the instantaneous rates. They are obtained by
differentiating the transition probability along the trajectory with respect to $\tau$,
which leads to \cite{schl,louk}
\bq
{\cal R}(E, \tau)=
2\,\mbox{Re}\int_{-\infty}^\tau d\tau^\prime
\, e^{-iE(\tau-\tau^\prime} D^+({\bf x}(\tau),{\bf x}(\tau^\prime))
\qquad\qquad\quad\,\,\,
\nonumber
\\
=2\, \mbox{Re}\int_{0}^\infty ds\, e^{-iEs} D^+({\bf x}(\tau),{\bf x}(\tau-s)),
\quad
s=\tau-\tau^\prime,
\label{intrat}
\eq
where we assumed as before that the interaction begins in the infinite past.
In the special case when the stationarity condition (\ref{invtra}) is obeyed
formula (\ref{intrat}) reduces to the time-independent rates (\ref{detrat}).

The rates in the problem of interest are in principle determined by (\ref{intrat}) together
with the Wightman function (\ref{wiggen}), in which we have to insert the trajectories
(\ref{tmp}) and (\ref{tdp}). The resulting expression is rather lengthy and we do not
write it down here. Unfortunately, an analytical calculation seems to be impossible in
this case, so we will be content with numerical results. Our main interest lies in making
clear the time evolution of the rates. Also, we will only focus on the excitation rates of
the detector, corresponding to transitions on positive energy levels $E>0$.

Extracting numerical values from (\ref{intrat}) is still not an immediate task and
a few preliminary observations are required. An important point concerns the regularization
of the Wightman function. It was remarked a few years ago by Schlicht \cite{schl} that, if one
tries to obtain the rates for the uniformly accelerated detector using (\ref{intrat}) with the
standard $i\varepsilon$ regularization, one does not recover the expected thermal rates
defined by (\ref{detrat}). The solution proposed in \cite{schl} is to adopt a different
regularization procedure, which essentially consists in assuming that the detector is an
extended object with a fixed dimension in its proper frame. The new regularization leads
indeed to the known thermal rates for uniform accelerations, as well as to physically
sensible results for various other trajectories \cite{barb, louk}.\footnote{Another nice
feature of the new regularization is that it restores the Lorentz invariance of the
regulated Wightman function: the fixed proper length of the detector is practically equivalent
to a frequency cutoff in the detector's proper frame, in contrast to the cutoff in the static
Minkowski frame in the usual regularization.} The analysis in \cite{schl} considered a massless
scalar field in four dimensions. In these conditions the Wightman function evaluated along an
arbitrary trajectory of the detector $\bf x(\tau)$ with the new regularization method has
the following form:
\bq
D^+({\bf x}(\tau), {\bf x}(\tau^\prime))=
\frac{1/4\pi^2}{({\bf x}(\tau)-{\bf x}(\tau^\prime)-i\varepsilon
(\dot {\bf x}(\tau)+\dot {\bf x}(\tau^\prime))^2}
\qquad\,\,\qquad \quad
\nonumber
\\
\nonumber
\\
=\frac{1/4\pi^2}{({\bf x}(\tau-i\varepsilon)-{\bf x}(\tau^\prime+i\varepsilon))^2},\
\quad\epsilon>0,
\quad \epsilon \rightarrow 0,\,\,\,\,\,
\label{wigreg}
\eq
where the squares denote the Minkowski norm ${\bf x}^2=-t^2+{\vec x}^{\,2}$ and the dot
represents derivation with respect to $\tau$. The dependence on $i\varepsilon$  in the
second fraction has to be interpreted in the sense of a formal Taylor expansion of the
functions $\bf x(\tau)$ in powers of $\varepsilon$, in which one retains the first two
terms.

Fortunately, there is no\, need to repeat the calculation in \cite{schl} for our
two-dimensional problem since the second formula in (\ref{wigreg})  allows a
straightforward generalization. The natural extension to an arbitrary Wightman function is
\bq
D^+({\bf x}(\tau), {\bf x}(\tau^\prime))=
D^+({\bf x}(\tau-i\varepsilon), {\bf x}(\tau^\prime+i\varepsilon)),
 \label{regpre}
\eq
where in the right member $D^+({\bf x}, {\bf x}^\prime)$ is the unregularized Wightman
function. In the case of interest this is simply given by (\ref{wiggen}) with $\varepsilon =0$.
The regularized form according to (\ref{regpre}) is obtained by making in the unregularized
quantity (\ref{wiggen}) the substitutions
\bq
(u, v)\rightarrow
(u(\tau - i\varepsilon),\, v(\tau - i\varepsilon)),
\quad
(u^\prime, v^\prime)\rightarrow
(u(\tau + i\varepsilon),\, v(\tau + i\varepsilon)),
\eq
with the same prescription for the dependence on $\varepsilon$. The rates\, shown in the plots
below are obtained from the numerical\, evaluation of (\ref{intrat}) with this regularization
method. (The value of $\varepsilon$ was decreased until the integrals remained practically
unchanged; for the parameters on the plots this happens for $\varepsilon \sim 10^{-5}$.)
More details on the calculation of the rates are given in Appendix B.

The evolution of the excitation rates\, for a number of energies $E$ is shown in Figs.~4.
The vanishing values before $\tau=0$ correspond to the interval $t<0$ in which both the
mirror and the detector are fixed. One sees that at sufficiently large times the rates become
constant. As expected, the values at $\tau\rightarrow \infty$ reproduce the thermal rates
for strictly infinite acceleration times (\ref{tramir}). The curves in Fig.~5 represent the
rates as a function of $E$ at different times $\tau>0$, showing the evolution towards
the purely thermal spectrum at infinite times. The KMS condition for the final rates can
be established from the fact that in this limit the Wightman function evaluated along the
detector's trajectory approaches the expression in Sec. 3.1, which can be easily checked with
numerical calculations (see Appendix B). It is interesting to observe from Fig.~5 that the
thermal spectrum is roughly attained after a proper time $\tau$ not very large (around ten units)
compared to the characteristic time defined by the inverse acceleration $1/a$.

\begin{figure}[H]

\center
\includegraphics[width=0.9\linewidth]{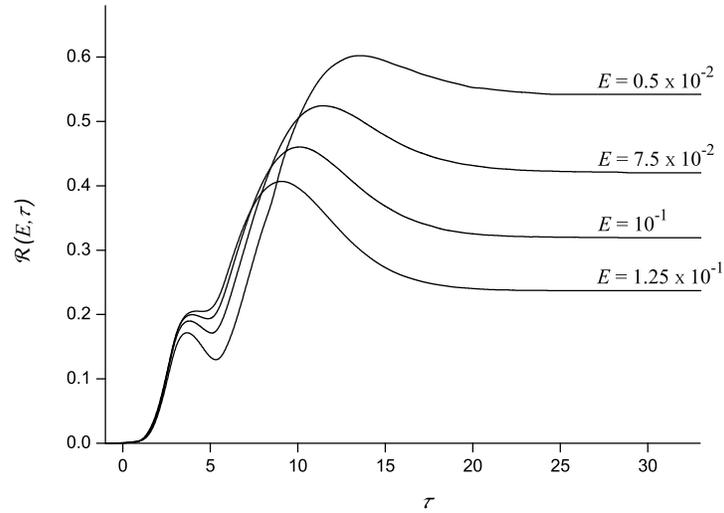}

\caption{The excitation rates of the detector as a function of the proper time
$\tau$ for different energies $E>0$. The mirror and the detector are static up
to the time $\tau=0$ after which they co-accelerate as indicated
in Sec. 4.1. The trajectory of the detector is defined by $\xi_D=1$. The differences
between the constant rates at late times and the thermal rates for infinite acceleration
times (\ref{tramir}) are invisible on the plot. The units for all quantities are
such that $a=1$ and the same for all the following figures.}

\end{figure}

\begin{figure}[H]

\center
\includegraphics[width=0.9\linewidth]{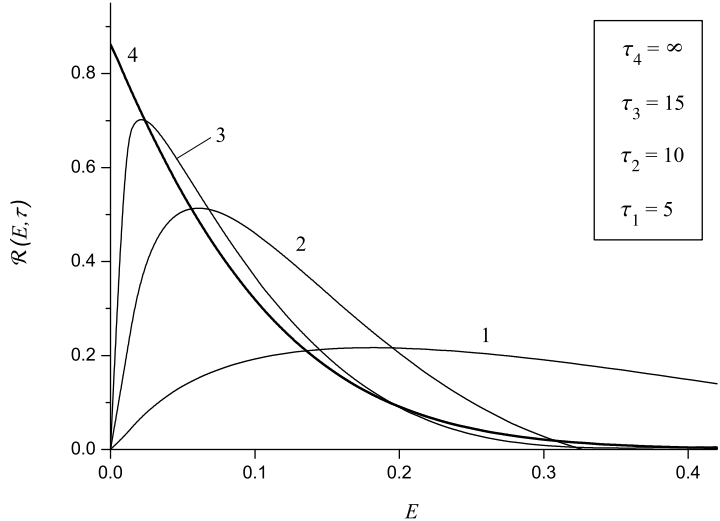}

\caption{The excitation spectrum of the detector at different times $\tau>0$ in the same
conditions as in Fig.~4. Curve $4$ (solid line) represents the purely thermal spectrum
for infinite acceleration times (\ref{tramir}).}

\end{figure}

Although not directly related to our subject, it is interesting to make at this point
a comparison with the rates for the same trajectory of the detector in the absence of
the mirror. In these conditions one has to use in (\ref{intrat}) the Wightman function for
the free field, which is given by the first logarithm in (\ref{wigacc}). It turns out that
this leads to a qualitatively different time dependence of the rates. The essential difference
is that the evolution acquires
a long-lasting oscillatory behaviour with $\tau$. Physically, this can be seen as a consequence
of the long range correlations for the massless field in two dimensions, which
produce in the transition amplitude a persistent interference effect between the contributions
due to the accelerated and the inertial part of the trajectory. In order to stabilize the rate,
one has to eliminate the second contribution. We have done this by evaluating (\ref{intrat}) with
a decoupling factor of the form $e^{-\eta s}$, with $\eta$ a small positive parameter. Figure 6
shows the evolution of the rates in these conditions. (Compare with
the times in Fig.~4.) The wild oscillations following the moment $\tau=0$ are typical
for a sudden variation of the kinematic parameters of the detector.\footnote{The relevance of
the low dimensionality of the problem for the oscillatory behaviour can be seen
from a comparison with the four dimensional examples in \cite{barb}. For essentially the same
trajectories of the detector the wild oscillations are absent and the evolution is very similar
to that in Fig.~4; see the plots in the cited paper.} Note that a larger decoupling parameter
$\eta$ eliminates the oscillations faster, but it introduces larger deviations from the expected
thermal rates at large acceleration times. A more detailed discussion on these points is given
in  Appendix B. A notable conclusion\, is the mechanism which explains the absence of the
long-lasting oscillations in the presence of the mirror: the Wightman function in this case
is given by the difference of two logarithms (like in (\ref{d0})-(\ref{d2})), and it so
happens that the oscillatory terms introduced by each logarithm cancel out in the difference.
Intuitively, one may say that the mirror destroys the long-range correlations of the field
along the detector's trajectory.

\begin{figure}[H]

\centering
\includegraphics[width=0.90\linewidth]{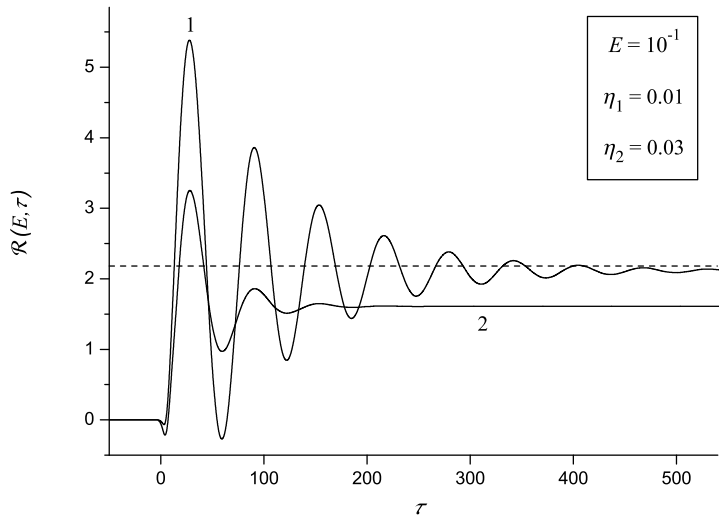}

\caption{Evolution of the excitation rates in the absence of the mirror for
two values of the decoupling parameter $\eta$ (see Appendix B) and the same
trajectory of the detector as in Fig.~4. The dashed line shows the thermal
rate in empty space (\ref{thra}) which corresponds to the uniform acceleration
after $\tau=0$.}

\end{figure}

{\large {\bf 4.2 The distribution of Rindler particles}}
\\
\\
We are interested in the beta coefficients defined by the new trajectories
(\ref{trainp})
%. We have to evaluate
\bq
\tilde \beta(\omega,\omega^\prime) =
(\tilde \varphi_\omega^{R\,*},\, \tilde \varphi_{\omega^\prime}^M),
\label{betcob}
\eq
where the scalar product is identical with that in (\ref{promir}) and the modes are given
by (\ref{minine}) and (\ref{rinine}). It is evident that for the Rindler modes the
decomposition (\ref{fg2}) remains valid, and as can be seen from (\ref{fominc})-(\ref{fomref2})
the same is true for the Minkowski decomposition (\ref{fg1}). As a consequence we can apply
the same calculation as in Sec. 3.2.

We begin with the following\, construction. Let us denote by $S$ the intersection
point between the hypersurface $\eta=0$ and the null ray reflected by the mirror at
point $J$ (see Fig.~3). We denote the spatial Rindler coordinate of this point
by $\xi_0$. One can see from the diagram that the null retarded coordinate of $S$
is identical to that of $J$, i.e. $u=u_0$. Using (\ref{nulrin}) one has that
\bq
\xi_0=\frac{1}{a} \ln (-a u_0).
\label{xis}
\eq
The relation with the time $t_0$ is
\bq
t_0=\frac{1}{2}(u_0+v_0)=-\frac{1}{a}\sinh(a\xi_0).
\eq
Note that the limit of infinite acceleration times is equivalent to
\bq
\lim_{t_0\rightarrow -\infty}
\xi_0=\infty.
\label{infati}
\eq

We now discuss the form of the Minkowski modes. Inserting (\ref{trainp})
in (\ref{minine}) and using (\ref{nulrin}) one finds following the notation
in (\ref{fg1})
\bq
F_\omega^{inc}(\xi+\eta)&=&\frac{1}{\sqrt{4\pi\omega}}\,
e^{-i (\omega/a)\,e^{a(\xi+\eta)}},
\label{fominc}
\eq
and
\bq
F_\omega^{ref}(\xi-\eta)&=&
\frac{1}{\sqrt{4\pi\omega}}\, e^{-i (\omega/a)\,e^{-a(\xi-\eta)}},
\quad \xi-\eta < \xi_0,
\label{fomref1}
\\
F_\omega^{ref}(\xi-\eta)&=&
\frac{1}{\sqrt{4\pi\omega}}\,
e^{-i (\omega/a)\,
\large [2\, e^{-a\xi_0}-e^{a(\xi-\eta)-2a\xi_0}\large]},
\quad \xi-\eta \geq \xi_0,
\label{fomref2}
\eq
where
we eliminated the parameter $u_0$ in favour of $\xi_0$ using (\ref{xis}). Note that
(\ref{fominc}) and the piece of the reflected component (\ref{fomref1}) are identical with
the corresponding quantities in the old modes (\ref{expmin}). The difference in the
new modes appears only in the reflected piece (\ref{fomref2}), which corresponds to the
reflection on the inertial part of the trajectory. Also notice that the definition domains
for (\ref{fomref1}) and (\ref{fomref2}) are delimited by the line $\xi-\eta=\xi_0$, or
equivalently $u=u_0$, which is the null ray reflected at point $J$.
\\
\\
{\bf 4.2.1 The beta coefficients}
\\
\\
Introducing (\ref{rinmod}) and (\ref{fominc})-(\ref{fomref2}) in (\ref{betcob})
one finds that the beta coefficients are (see Appendix A)
\bq
\tilde \beta(\omega,\omega^\prime)=
-\frac{1}{2\pi} \sqrt {\frac{\omega}{\omega^\prime}}
\left \{b_1 (\omega,\omega^\prime)+ b_{2} (\omega,\omega^\prime) \right\},
\label{newbet}
\eq
where
\bq
b_1 (\omega,\omega^\prime)&=&
\int_{-\xi_0}^\infty d\xi\,
e^{-i\omega \xi -i (\omega^\prime/ a)\, e^{a\xi}},
\label{B1}
\\
b_{2} (\omega,\omega^\prime)&=&
\int^{-\xi_0}_{-\infty} d\xi\, e^{-i\omega \xi-i (\omega^\prime/a)\,
\large [2\, e^{-a\xi_0}-e^{-a(\xi+2\xi_0)}\large]}.
\label{B2}
\eq
Before considering in more detail (\ref{newbet})-(\ref{B2}) a number of observations are
necessary. It is quite evident that the new Bogolubov coefficients\, do not reproduce the
old coefficients (\ref{mircoe}). Given the unproblematic nature of the new trajectories,
one can expect that the new coefficients will preserve the purity of the initial vacuum
in terms of Rindler states. For example, a significant difference between the two sets of
coefficients will manifest in expectations values in the initial vacuum of the form\,
$\langle a_{\omega_1}^{R+} a^R_{\omega_2}\rangle$, where $a^R_\omega$ and $a_{\omega}^{R+}$
are the annihilation and creation for the Rindler particles. Since a thermal statistical
matrix is diagonal in the particle basis, the purely thermal coefficients in Sec. 3.2 imply
an identically zero result when $\omega_1\neq \omega_2$. As we will see in Sec. 4.3.2, this
is not the case for the new coefficients.

Another essential point concerns the limit of infinite acceleration times $t_0\rightarrow -\infty$,
or equivalently $\xi_0\rightarrow \infty$. Note that the old coefficients (\ref{mircoe}) can be
recovered from (\ref{newbet}) by setting\, $\xi_0=\infty$ in the first integral (\ref{B1}) and
ignoring the second integral (\ref{B2}). From the integration limits in the two integrals, one would
expect that for $\xi_0\rightarrow \infty$ the new coefficients will reduce to the old ones. We will
show below that this is not so. As we have remarked in Sec. 1, this can be seen to reflect the fact
that the purity of the initial vacuum is unaffected by the choice of $t_0$.

We now establish a more convenient form of the coefficients for extracting the limit
$\xi_0\rightarrow \infty$. Introducing the complex variables $z_1$ and $z_2$ defined
by
\bq
b_1: \,\,\,
z_1=\frac{\omega^\prime}{a}\, e^{a\xi}-\frac{\omega^\prime}{a}\, e^{-a\xi_0},
\quad\,\,
b_{2}:\,\,\,
z_2=
\frac{\omega^\prime}{a}\, e^{-a(\xi+2\xi_0)}-\frac{\omega^\prime}{a}\, e^{-a\xi_0},
\eq
one can rewrite (\ref{B1}) and (\ref{B2}) using a contour integration in the complex
plane\footnote{We rotate the real integration axis corresponding to $z_1, z_2\in [0, \infty)$
in the original integrals along the negative (positive) imaginary axis in the first (second)
integral.} in the following way (we ignore a common irrelevant phase factor):
\bq
b_1 (\omega,\omega^\prime)=
-\frac{i}{a}
\left (\frac{a}{\omega^\prime}\right)^{-i\omega/a}
\int_0^\infty dt_1
\left (\frac{\omega^\prime}{a}\, e^{-a\xi_0} -it_1\right)
^{-i\omega/a-1} e^{-t_1},
\label{bo1}
\\
b_{2} (\omega,\omega^\prime)=e^{2i\omega \xi_0}\times
\frac{i}{a} \left (\frac{a}{\omega^\prime}\right)^{i\omega/a}
\int_0^\infty dt_2
\left (\frac{\omega^\prime}{a}\, e^{-a\xi_0} +it_2\right)^{i\omega/a-1} e^{-t_2}.
\label{bo2}
\eq
Letting $\xi_0\rightarrow \infty$ in the integrands the factors $e^{-a \xi_0}$ in the
large parentheses vanish and one is tempted to express the integrals in terms of the
Gamma functions $\Gamma(z=\mp i \omega/a)$. However, this is not justified due to the
convergence condition Re\,$z>0$, which is not respected here. The problem can be fixed
by performing an integration by parts writing first the parentheses as a derivative,
which increases the arguments of the Gamma functions by one. The result is
\bq
b_1 (\omega,\omega^\prime)=-\frac{i}{\omega}\, e^{i \omega \xi_0}
+\frac{i}{\omega}
\left (\frac{a}{\omega^\prime}\right)^{-i\omega/a}
\int_0^\infty dt_1
\left (\frac{\omega^\prime}{a}\, e^{-a\xi_0} -it_1\right)
^{-i\omega/a} e^{-t_1},
\label{bb1}
\\
b_{2} (\omega,\omega^\prime)=
+\frac{i}{\omega}\, e^{i \omega \xi_0}
\qquad\qquad\qquad\qquad\qquad\qquad\qquad\quad\quad\qquad\qquad\quad\,\,\,
\nonumber
\\
-e^{2i\omega \xi_0}\times \frac{i}{\omega}
\left (\frac{a}{\omega^\prime}\right)^{i\omega/a}
\int_0^\infty dt_2
\left (\frac{\omega^\prime}{a}\, e^{-a\xi_0} +it_2\right)
^{i\omega/a} e^{-t_2}.
\qquad
\label{bb2}
\eq
The problematic limit $\xi_0\rightarrow \infty$ reappeared in the oscillatory
boundary terms $\sim \mp i e^{i \omega \xi_0}$, but the integrals are well-defined.
It is nice that in the sum (\ref{newbet}) these terms cancel each other out.

The limit can now be easily obtained from (\ref{bb1}) and (\ref{bb2}) by letting
$\xi_0\rightarrow \infty$ under the integrals, which leads to
\bq
\lim_{\xi_0\rightarrow \infty}
\tilde \beta(\omega, \omega^\prime)
=-\frac{1}{2\pi} \sqrt {\frac{\omega}{\omega^\prime}}
\left \{B_1 (\omega,\omega^\prime)+B_{2} (\omega,\omega^\prime)\right\},
\label{limb12}
\eq
where
\bq
B_1 (\omega,\omega^\prime)
&=&
\frac{i}{\omega}
\left (\frac{a}{\omega^\prime}\right)^{-i\omega/a}
e^{-\pi \omega/2a}\,
\Gamma(-i\omega/a +1),
\label{BI}
\\
B_{2} (\omega,\omega^\prime)
&=&- e^{2i\omega \xi_0}
\times
\frac{i}{\omega}
\left (\frac{a}{\omega^\prime}\right)^{i\omega/a}
e^{-\pi \omega/2a}\,
\Gamma(+i\omega/a +1).
\label{BII}
\eq
Note that the dependence on\, $\xi_0$ survives in (\ref{BII}). Using in (\ref{BI}) the recurrence
relation
\bq
\Gamma(-i\omega/a+1)=(-i\omega/a)\Gamma(-i\omega/a),
\eq
one finds that the term $B_1$ in (\ref{limb12}) reproduces the old coefficients (\ref{frecoe}), i.e.
\bq
\bar \beta(\omega, \omega^\prime)=
-\frac{1}{2\pi} \sqrt {\frac{\omega}{\omega^\prime}}
B_1 (\omega,\omega^\prime).
\eq
It is evident that the term $B_2$ does not vanish, so that it is clear that for $\xi_0\rightarrow \infty$
the new coefficients do not reduce to the old coefficients.

Let us point out that in the calculation of the old/empty space coefficients, which are defined only
by (\ref{B1}) with $\xi_0\rightarrow \infty$, the integration by parts in (\ref{bo1}) is not necessary.
In this case the integral can be made convergent by simply adding to the exponent $-i\omega/a-1$ a
small quantity $\lambda >0$, from which (\ref{frecoe}) follows with $\lambda\rightarrow 0$.
The same procedure is inconvenient for the $\xi_0$-dependent coefficients, since it turns out
that this mixes the dependence on the regulator $\lambda$ with that on $\xi_0$, which is undesirable.
For example, repeating the integration by parts with an extra term $\lambda>0$ in the exponents in
(\ref{bo1}) and (\ref{bo2}) one finds that the oscillatory boundary terms in (\ref{bb1}) and (\ref{bb2})
do not cancel among themselves, which introduces artificial oscillations with $\xi_0$. In addition,
these terms acquire a factor $e^{-a \lambda \xi_0}$, which leads to ambiguities when trying to remove
the regulator $\lambda \rightarrow 0$ and simultaneously let $\xi_0\rightarrow \infty$.
\\
\\
{\bf 4.3.2. Emergence of the thermal statistics for $t_0\rightarrow -\infty$}
\\
\\
We have seen that for $\xi_0\rightarrow \infty$ the term $B_1$ in (\ref{limb12}) reproduces the
empty space coefficients (\ref{frecoe}). Hence, in order to recover the thermal statistics
of the Rindler particles described by these coefficients it is sufficient to get rid of the term $B_{2}$.
The mechanism by which one can do that is rather immediate from the oscillatory factor $e^{2i\omega \xi_0}$
in (\ref{BII}). The basic observation is that, in a physically realistic situation, one does not measures
quantities associated to a precise Rindler frequency $\omega$, but to finite norm states with a non-zero
frequency width $\Delta \omega>0$. This means that the relevant Bogolubov coefficients are obtained
with a smearing with respect to the frequencies $\omega$. It becomes then clear that after integration
over $\omega$ the highly oscillatory factor $e^{2i\omega \xi_0}$ for $\xi_0\rightarrow \infty$
will eliminate the non-thermal  term  $B_{2}$. Thanks to relation (\ref{alfbet}) the same
mechanism applies to the alpha coefficients. This practically provides the desired connection
with the thermal statistics obtained in the first calculation.

To be more precise, let us denote by $f_{\Omega}(\omega)$ the smearing functions associated to some set
of finite norm Rindler states, with $\Omega$ standing for the mean frequency in the wave packet, and let
us introduce the smeared annihilation and creation Rindler operators
\bq
\tilde a_{\,\Omega}^R=\int d\omega\, f_\Omega(\omega)\, \tilde a_{\,\omega}^R,
\quad
\tilde a_{\,\Omega}^{R\,+}=\int d\omega\, f^*_\Omega(\omega)\,
\tilde a_{\,\omega}^{R\,+}.
\label{smeope}
\eq
Then\, the effective Bogolubov coefficients which define the measurable expectation
values are
\bq
\tilde
\alpha(\Omega, \omega^\prime)=
\int d\omega\, f(\omega)\,
\tilde \alpha(\omega, \omega^\prime),
\qquad
\tilde \beta(\Omega, \omega^\prime)=
\int d\omega\, f^*(\omega)\,
\tilde \beta(\omega, \omega^\prime).
\label{smecoe}
\eq
Relations identical with (\ref{smeope}) and (\ref{smecoe}) can be written for the quantities in
Sec.~3.2 with the tilde replaced by the bar. According to the argument above, the emergence of
the effective thermal statistics from the initial vacuum is expressed by
\bq
\lim_{\xi_0\rightarrow \infty}
\tilde \alpha(\Omega, \omega^\prime)
= \bar \alpha(\Omega, \omega^\prime),
\quad
\lim_{\xi_0\rightarrow \infty}
\tilde \beta(\Omega, \omega^\prime)
= \bar \beta(\Omega, \omega^\prime).
\label{terpro}
\eq

In order to illustrate the ``thermalization process'' implied by (\ref{terpro}) with some  explicit
examples, let us focus on expectation values of the form
\bq
n(\Omega_1, \Omega_2)
\equiv
\langle \tilde a_{\,\Omega_1}^{R\,+}\, \tilde a_{\,\Omega_2}^R\rangle.
\label{ngen}
\eq
Using (\ref{smecoe}) together with the standard definitions of the Bogolubov coefficients in terms of
the creation and annihilation operators \cite{birr, cris} one has that
\bq
n(\Omega_1, \Omega_2)
\qquad\qquad\qquad\qquad\qquad\qquad\qquad\qquad\qquad
\qquad\qquad\qquad\qquad
\nonumber
\\
=\int_{\Delta \omega_1} d\omega_1 \int_{\Delta \omega_2} d\omega_2 \int_0^\infty d\omega^\prime
f^*_{\Omega_1}(\omega_1) f_{\Omega_2}(\omega_2)\,
\tilde \beta (\omega_1, \omega^\prime) \tilde \beta^*(\omega_2, \omega^\prime),
\label{numsme}
\eq
where we introduced  $\Delta \omega_{1,2}$ the frequency windows of the smearing functions. We recall
that in empty space the thermal
expectation values defined by the (unsmeared) beta coefficients (\ref{frecoe})
are \cite{taka1}
\bq
\langle a_{\,\omega_1}^{R\,+} a_{\,\omega_2}^R
\rangle =
\int_0^\infty d\omega^\prime
\beta (\omega_1, \omega^\prime) \beta^*(\omega_2, \omega^\prime)
\nonumber
\\
=\delta(\omega_1-\omega_2) \frac{1}{e^{2\pi \omega_1/a}-1}.
\quad\,\,
\label{delbet}
\eq
As a consequence of (\ref{idenb}) the same expectation values are valid for the barred coefficients
in Sec.~3.2. The second limit in (\ref{terpro}) then implies
\bq
\lim_{\xi_0\rightarrow \infty}
n(\Omega_1, \Omega_2)
\qquad\qquad\qquad\qquad\qquad\qquad\qquad\qquad\qquad\qquad
\qquad
\nonumber
\\
\qquad\qquad\quad
= \int_{\Delta \omega_1}
d\omega_1
 \int_{\Delta \omega_2}
d\omega_2\,
\delta(\omega_1-\omega_2) f^*_{\Omega_1}(\omega_1) f_{\Omega_2}(\omega_2)
\frac{1}{e^{2\pi \omega_1/a}-1}.
\label{smeter}
\eq

Unfortunately, due to the complicated form of our coefficients\, an analytical calculation for
(\ref{numsme}) is virtually impossible, so we will again resort to numerical results. A direct
numerical calculation based on (\ref{numsme}) however is not simple, one major problem being the
highly oscillatory factors $e^{2i\omega\xi_0}$ when $\xi_0\rightarrow \infty$. A natural choice
in order to simplify the integral is to assume very small frequency widths $\Delta \omega$,
which allows to use Taylor expansions with respect to $\omega_{1,2}$ in the integrand. An
approximation along this line which reduces (\ref{numsme}) to a one-dimensional integral is
presented in Appendix C. The smearing functions $f_{\Omega}(\omega)$ used in the calculation
are unit-norm functions of rectangular shape centered in $\Omega$ with $\Delta \omega \ll \Omega,\,
a$ (see (\ref{smefun})). The plots below are based on the  integrals obtained there.

In Figures 7-10\, we represented $n(\Omega_1, \Omega_2)$ as a function of $\xi_0$ for different
values of $\Omega$ and $\Delta \omega$. Note that for identical smearing functions
$\Omega_1=\Omega_2\equiv \Omega$ the expectation values (\ref{numsme}) are just the mean numbers
of Rindler particles defined by the wave packet $f_\Omega(\omega)$ in the initial vacuum,
\bq
n(\Omega)=\langle \tilde a_{\,\Omega}^{R\,+} \tilde a_{\,\Omega}^R\rangle.
\label{numrin}
\eq
This case is represented in Figs. 7 and 8. One sees that for $\xi_0$ sufficiently large
the particle numbers approach the thermal values defined by (\ref{smeter}),
i.e.
\bq
\lim_{\xi_0\rightarrow \infty} n(\Omega)
=
\int_{\Delta \omega}
d\omega\,
\vert f_{\Omega}(\omega)\vert^2
\frac{1}{e^{2\pi \omega/a} -1}
\simeq\frac{1}{e^{2\pi \Omega/a} -1}.
\label{ilius}
\eq
The thermal numbers are not exactly recovered as a consequence of our approximations
(note, however, the values on the $y$-axis). It is interesting that already at $\xi_0=0$
(which corresponds to $t_0=0$) the particle numbers are very close to the thermal values.
Figures 9 and 10 illustrate the case $\Omega_1\neq \Omega_2$. We have chosen the parameters
in such a way that the frequency windows do not overlap, i.e. $\Omega_1-\Omega_2>\Delta \omega$.
It is immediate from (\ref{smeter}) that in these conditions
\bq
\lim_{\xi_0\rightarrow \infty} n(\Omega_1, \Omega_2)=0.
\label{vanbe}
\eq
The vanishing behaviour of the expectation values when $\xi_0\rightarrow \infty$ can be seen
on the plots. As one could have anticipated, a larger difference $\Omega_1-\Omega_2$ implies
a faster approach to zero.

\begin{figure}[H]

\centering

\includegraphics[width=0.9\linewidth]{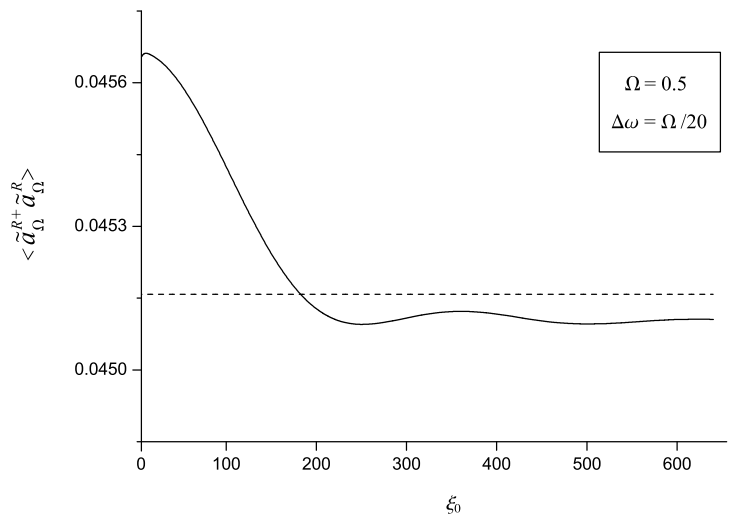}

\caption{The number of Rindler particles in the initial vacuum $n(\Omega)$ defined by a given
$f_\Omega$ evaluated according to Appendix C represented as a function of $\xi_0$. The parameters
of the smearing function are shown in the box. The dotted line indicates the exact value defined
by the effective thermal statistics (\ref{smeter}).}

\end{figure}

\begin{figure}[H]

\centering

\includegraphics[width=0.9\linewidth]{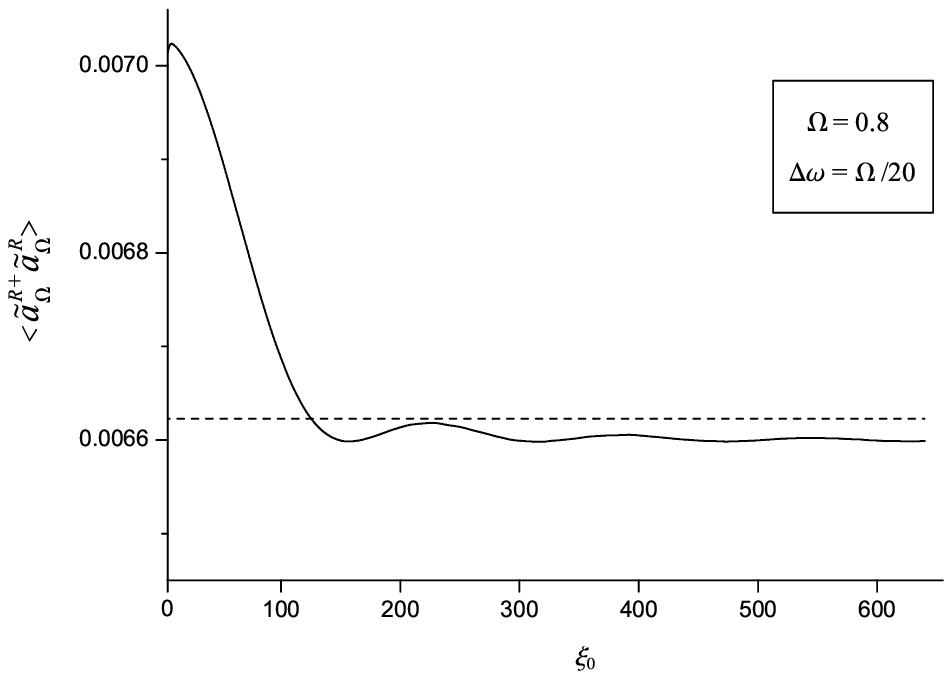}

\caption{The same as in Fig.~7 for different parameters of the smearing function.}

\end{figure}

\begin{figure}[H]

\centering

\includegraphics[width=0.9\linewidth]{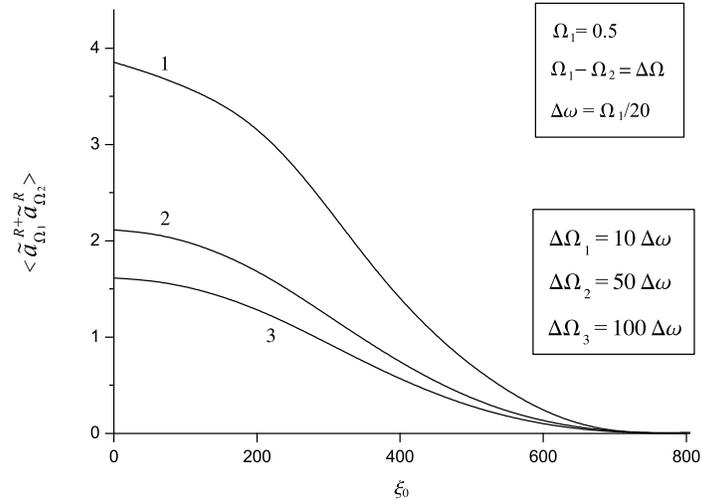}

\caption{The same as in Fig.~7 but for the expectation values $n(\Omega_1, \Omega_2)$
defined by non-identical smearing functions $f_{\Omega_1}\neq f_{\Omega_2}$. The parameters
are such that $\Omega_1-\Omega_2>\Delta \omega$, so that the smearing functions do not overlap.
In these conditions the expectation values in a thermal state of the field are exactly zero.}

\end{figure}

\begin{figure}[H]

\centering

\includegraphics[width=0.9\linewidth]{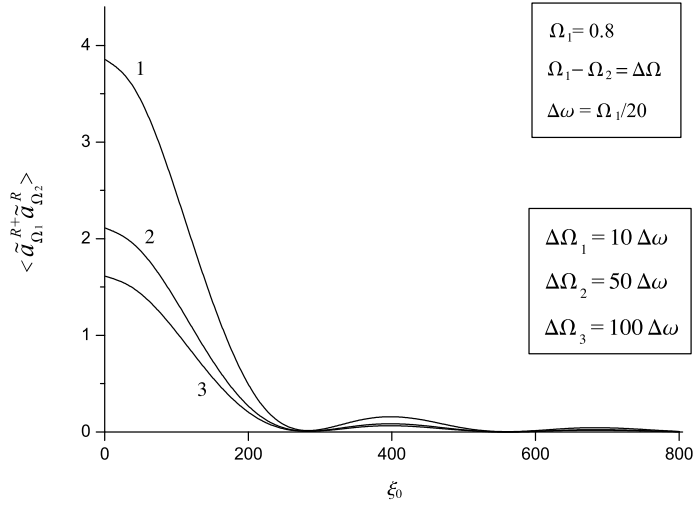}

\caption{The same as in Fig.~9 for different parameters of the smearing functions.}

\end{figure}

{\bf 4.3.4 An analogy with Unruh effect in empty space}
\\
\\
As a final point, let us present an intuitive explanation for the emergence of the
effective thermal statistics established above. In the usual discussion of the Unruh effect,
the fact that the vacuum appears as a mixed state for the Rindler observers can be understood
in terms of the unobservable degrees of freedom of the field masked by the Rindler horizon.
One can then ask what explains the mixed state in our case, since in the presence of the mirror
no unobservable degrees of freedom exist. The unsmeared coefficients $\tilde \beta (\omega,
\omega^\prime)$ which do not reduce to the purely thermal coefficients even for
$t_0\rightarrow -\infty$ can be seen to be a reflection of this fact. However, it is
not hard to guess that the picture changes when we restrict to the expectation values of
smeared Rindler operators. We will now explain that this effectively introduce unobservable
degrees of freedom in the system, which leads to a picture very similar to that for the effect
in the empty space. We stress that the discussion which follows is only a qualitative one.

The argument is as follows. We consider again Fig.~3. The first observation has to do with
the state of the field on the hypersurface $\eta=0$, denoted by $\Sigma$. We recall that
point $S$ is the intersection of this hypersurface and the last null ray reflected on the
inertial part of the trajectory and that the spatial Rindler coordinate of this point
is $\xi_0$ (see (\ref{xis})). One can then read from the null rays that: (1) in the region
$0<\xi<\xi_0$ the state of the field is locally identical with the perturbed vacuum in the
presence of the infinitely accelerated mirror in Sec.~3.2, and (2) in the region $\xi_0<\xi<\infty$
the state of the field is locally identical with the vacuum in presence of the
inertial mirror. By ``locally identical'' we mean that within each region all correlation
functions coincide, which follows from the fact that
within these regions the Minkowski modes
that describe the two states coincide.\footnote{In a non-interacting theory all correlation
functions are completely defined by the free modes.} Also, note that the purely thermal
vacuum in Sec.~3.2 corresponds to the case when the local vacuum zone $\xi>\xi_0$  is
completely absent. This strongly suggest that, in order to account for the  purity of the
initial vacuum in terms of the state of the field on $\Sigma$, it is crucial to pay attention
to the degrees of freedom in this region.

The second observation is that for a smearing function of finite norm in the $\omega$-space
the Fourier transformed wave packet in the position $\xi$-space automatically vanishes for
$\xi\rightarrow \infty$. The important implication is that operators smeared with such
functions will practically ignore the field at sufficiently large Rindler distances $\xi$.
As a consequence, these degrees of freedom become now the ``unobserved'' degrees of freedom
of the system. Recalling the essential role of the degrees of freedom at large $\xi$ for
preserving the purity of the initial vacuum noted above, this practically explains the
emergence of the mixed state in our problem. The limit $\xi_0\rightarrow \infty$ in which we
recovered the thermal statistics was needed in order to ensure that the essential vacuum
zone $\xi>\xi_0$ is pushed sufficiently far away in the region in which the smearing functions
vanish for an arbitrary form of these functions. We are thus led to a picture very similar to
that for the Unruh effect in empty space: the degrees of freedom of the field in the unobserved
Rindler wedge correspond here to the degrees of freedom at large Rindler distances from the mirror,
combined with the limit of large acceleration times $t_0\rightarrow -\infty$. It deserves to be
remarked that the picture we have skecthed here allows to conclude that the thermal expectation
values of the smeated operators can be also obtained for a finite $t_0$ (finite $\xi_0$), as
long as we make sure that the support of the smearing functions in the $\xi$-space falls completely
outside the local vacuum region $\xi>\xi_0$.

\section*{5. Conclusions}

Let us summarize our main results. We investigated the Unruh effect in a two
dimensional Minkowski space in the presence of a uniformly  accelerated perfect
mirror, considering the space-time region for which the Rindler horizon is masked
by the mirror, so that no unobserved degrees of freedom of the system exist. We
found that the characteristic thermal properties of the effect survive, i.e.
the response of a co-accelerated particle detector and the distribution of the
Rindler particles in the initial vacuum still display a thermal form. Our
result thus adds further evidence in favour of the views which emphasize the local
nature of the effect, along with its independence from the Rindler
horizon \cite{niko, rove, rava1, rava2}.

For the trajectories of the mirror with accelerations which start at $t\rightarrow -\infty$,
the transition rates of the co-accelerated detector are exactly
thermal, in the sense that the KMS condition is obeyed. The distribution of the Rindler
particles is also exactly thermal, being essentially identical to that in empty space.
The last conclusion, however, is inconsistent with the absence of the unobserved
degrees of freedom of the field in the presence of the mirror and the purity of
an initial vacuum. We ascribed this inconsistency to the non-vanishing accelerations
in the infinite past, which led us to consider the case when the mirror starts to
accelerate at a finite time $t_0$.

For the new trajectories we found that the rates of a comoving detector at sufficiently
large times approach, as expected, the thermal rates in the previous calculation. An
interesting phenomenon emerged at a comparison with the rates in the absence of the
mirror, in which conditions for the same trajectories of the detector the rates show
a pronounced long-lasting oscillatory behaviour, whose origin can be traced to the
long-range correlations for the massless field in two dimensions. It appears that
the mirror eliminates this phenomenon.

A notable point concerning the Bogolubov coefficients for the new trajectories is that for
$t_0\rightarrow -\infty$ they do not reduce to the thermal coefficients in the previous
case. We have interpreted this as a reflection of the fact that the new coefficients
preserve the purity of the initial vacuum state in terms of Rindler states, irrespective
of the choice of $t_0$. Nevertheless, we have shown that an effective thermal statistics can be
recovered for $t_0\rightarrow -\infty$ provided one restricts to measurements of operators
constructed from smeared creation and annihilation Rindler operators associated
to finite norm states. Formally, the mechanism is that the non-thermal terms
in the Bogolubov coefficients contain a factor which in the limit becomes highly oscillatory
with the Rindler frequencies, which eliminates these terms from the smeared quantities.
From a more physical point of view, the emergence of the thermal statistics is explained by
the fact that the smeared operators collect information only from finite Rindler distances
from the mirror. As a consequence of the special trajectory of the mirror, the state of
the field within any such region for $t_0\rightarrow -\infty$ becomes identical with the
purely thermalized vacuum for infinite acceleration times, which explains the result.

Finally, let us mention a possible extension of our investigation. It would be interesting to
give a more detailed description of the quantum state of the field on the hypersurface $\eta=0$, along
with the progressive thermalization for $t_0\rightarrow -\infty$. A natural way to do this is
by using the concept of entanglement entropy, which is generally defined as the von Neumann
entropy of the reduced density matrix associated to a part of the system, see e.g. \cite{benn}.
In our case a quantity of immediate interest would be the entanglement entropy associated
to the thermal region $\xi<\xi_0$, or, equivalently, the entropy due to ignoring\, the vacuum
zone $\xi>\xi_0$. At first sight, one would expect to recover the Gibbs entropy of a thermal
gas in Rindler space in the box $0\leq \xi\leq \xi_0$. We want to point out that existing
calculations in similar situations in which one restricts to a part of a thermal system show
that the two quantities\, generally agree only in the limit of large temperatures\, and/or large
dimensions of the box \cite{cala1, cala2}. This indicates that the entanglement entropy in our
problem will reproduce the Gibbs entropy only in the limit of large accelerations of the mirror
and/or large acceleration times, which probably will appear as the condition $a\xi_0\gg 1$.
We leave this as a subject for further research.

\section*{Acknowledgements}

I thank my friend Attila Farkas for the many discussions on the paper. The result for the 
detector's response in the presence of the co-accelerated mirror (\ref{tramir}), (\ref{frofac}) 
is directly inspired by his calculations.

\section*{Appendix A}

\setcounter{equation}{0}

\renewcommand{\theequation}{A.\arabic{equation}}

We arrive here to the general formula for the beta coefficients (\ref{genint}). We insert
(\ref{fg1})-(\ref{deri2}) in the integral (\ref{promir}) with $\eta=0$, which leads to
\bq
\bar\beta(\omega, \omega^\prime)=
i\int_0^\infty d\xi\, (G_\omega^{inc}-G^{ref}_\omega)
({F_{\omega^\prime}^{inc}}^{\,\prime} + {F_{\omega^\prime}^{ref}}^{\,\prime})
\quad
\nonumber
\\
-\omega \times
\int_0^\infty
d\xi\, (G_\omega^{inc}-G^{ref}_\omega) (F_{\omega^\prime}^{inc}-
F_{\omega^\prime}^{ref}).
\,\,\,
\label{a1}
\eq
In\, the first integral we eliminate the derivatives of $F_{\omega^\prime}$  with an
integration by parts and replace the resulting derivatives of $G_{\omega}$ using
(\ref{deri2}). The result is
\bq
i\int_0^\infty d\xi\,(\dots) =i(G_\omega^{inc}-G^{ref}_\omega)
(F_{\omega^\prime}^{inc} + F_{\omega^\prime}^{ref}) \Big\vert_{0}^\infty
\qquad\quad\,\,\,\,
\nonumber
\\
-\omega \times \int_0^\infty d\xi\,
(G_\omega^{inc}-G^{ref}_\omega)
(F_{\omega^\prime}^{inc}+F_{\omega^\prime}^{\,ref}).
\label{a2}
\eq
The boundary terms in (\ref{a2}) are zero due to the vanishing of the Rindler modes (the first
parenthesis) on the mirror $\xi=0$ and with the usual assumption that the field vanishes at
infinite distances $\xi\rightarrow \infty$. Adding the integral term in (\ref{a2}) to the
second integral in (\ref{a1}) one obtains (\ref{genint}).

We now refer to the coefficients for the trajectories in Sec. 4. It is clear that the calculation
above does not depend on the specific form of $F_\omega$ and $G_\omega$, which means that
(\ref{genint}) can be applied to these trajectories too. The new functions $F_\omega$ are defined
by (\ref{fominc})-(\ref{fomref2}), while  $G_\omega$ remain unchanged. Applying (\ref{genint}) one
finds
\bq
\tilde \beta(\omega,\omega^\prime) =
-\frac{1}{2\pi} \sqrt {\frac{\omega}{\omega^\prime}}
\left\{
\int_0^\infty d\xi\,e^{-i\omega \xi -i (\omega^\prime/ a)\,e^{a\xi}}
+\int_0^{\xi_0} d\xi\, e^{i\omega \xi -i(\omega^\prime/ a)\, e^{-a\xi}}
\right\}
\nonumber
\\
\nonumber
\\
-\frac{1}{2\pi} \sqrt {\frac{\omega}{\omega^\prime}}
\int_{\xi_0}^\infty d\xi\, e^{i\omega \xi-i (\omega^\prime/a)\,
\large [2\, e^{-a\xi_0}-e^{a(\xi-2\xi_0)}\large]}.
\qquad\qquad\qquad\,\,\,\,
\label{b1}
\eq
Note that (\ref{b1}) differs from (\ref{mircoe}) only via the third term which contains
the reflected Minkowski component for $\xi>\xi_0$. Integrating in the last two terms
with respect to $\xi\rightarrow -\xi$ and summing the first two integrals into a single
integral one arrives to (\ref{newbet})-(\ref{B2}).

\section*{Appendix B}

\setcounter{equation}{0}

\renewcommand{\theequation}{B.\arabic{equation}}

We detail here the evaluation of the rates (\ref{intrat}). We first make precise the
formula used in numerical calculations. We refer throughout the appendix to the
trajectories inertial in the past (\ref{tmp}) and (\ref{tdp}). It is convenient
to begin with the case of the detector in the absence of the mirror. In these conditions
the Wightman function is given only by the first logarithm in (\ref{wigacc}), which
implies that $D^+ ({\bf x}(\tau),\, {\bf x}(\tau-s))\sim \ln s$ for large $s$, so that
(\ref{intrat}) is ill-defined. The convergence of the integral can be assured by introducing
an adiabatic decoupling of the interaction in the infinite past via a factor $e^{-\eta s}$ with
$\eta>0$, which allows to perform an integration by parts, which makes the integrand $\sim 1/s$.
With these modifications the integral for the rates becomes
\bq
{\cal R}(E, \tau)=\frac{2}{E}\,\mbox{Im}
\int_0^\infty ds\, e^{-i Es -\eta s}\, W(\tau, s),
\label{newint}
\\
W(\tau, s)\equiv\frac{\partial}{\partial s}\,D^+ ({\bf x}(\tau),\, {\bf x}(\tau-s)).
\qquad\quad\,\,
\label{derwig}
\eq
We ignored in (\ref{newint}) the divergent boundary term due to the coincidence limit $s\rightarrow 0$,
which can be justified by the fact that for the massless field in two dimensions the Wightman function
is defined up to an additive constant, which can be chosen to eliminate this term. The plots presented
in Sec.~4.1 are based on the numerical evaluation of (\ref{newint}) with $D^+({\bf x},\, {\bf x}^\prime)$
regularized as discussed in the text. For the free field and uniform accelerations the formula
above with $\eta=0$ leads as it should to the standard thermal rates (\ref{thra}).

The essential observations for the detector in the absence of the mirror are: (1) for large $s$ one
has $W(\tau, s)\sim 1/s$, which makes (\ref{newint}) convergent even for $\eta=0$, and (2) the integral
with $\eta=0$ acquires a term that indefinitely oscillates with $\tau$. The source of this term can be
understood from Fig.~11, which shows\footnote{Only the real part is represented. The imaginary part
becomes negligible for $\varepsilon$ sufficiently small. The decoupling factor $e^{-\eta s}$ is not
included on the plot.} $W(\tau, s)$ as a function of $s=\tau-\tau^\prime$\, for different values of\, $\tau$.
The various pieces of the curves have the following origin: (1) the large values near $s=0$ correspond to
the coincidence limit $\tau^\prime \rightarrow \tau$, (2) the horizontal pieces result from $\tau^\prime$
in the acceleration phase of the trajectory with $a_D s\gg 1$ (the Wightman function in this case can be
approximated with (\ref{d1}) where $\ln \sinh^2 (a_Ds/2)\simeq a_D s$, and thus $W(\tau, s)\simeq\,$
constant), and (3) the tails at large $s$ result from the inertial phase $\tau^\prime<0$. The oscillatory
term in (\ref{newint}) is produced by the factor $e^{-i Es}$ integrated over the intervals which
correspond to the horizontal pieces and it is directly related to the fact that the area below the
curves indefinitely increases with $\tau$. The role of the decoupling factor $e^{-\eta s }$ is to
eliminate this contribution.

\begin{figure}[H]

\centering

\includegraphics[width=0.9\linewidth]{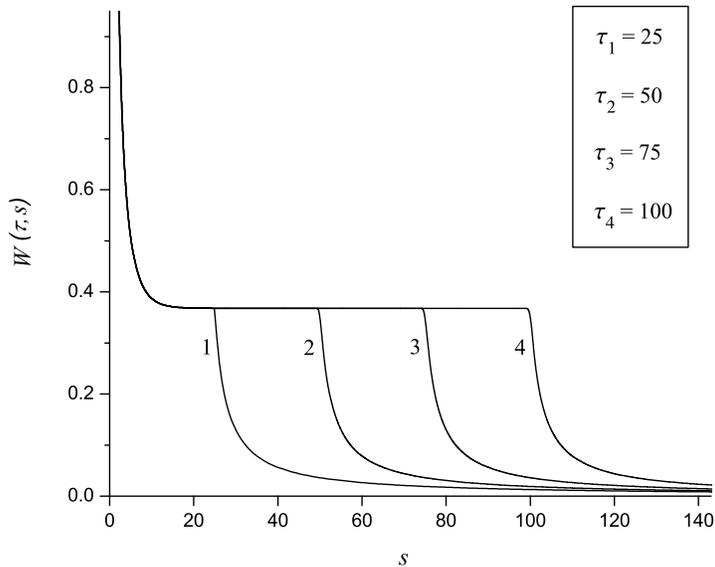}

\caption{$W(\tau, s)$ in the absence of the mirror represented as a function of $s$
for different times $\tau$. The trajectory of the detector is the same as in Fig.~4.
The long-lasting oscillatory behaviour of the rates in Fig.~6 results from the
contribution of the horizontal pieces of the curves.}

\end{figure}

A different picture arises in the presence of the mirror. The analogous curves are shown
in Fig.~12. Note that the curves have now two peaks. The peak at $s=0$ is the same with
that in Fig.~11, while the extra peak expresses the large correlations between points
${\bf x}$, ${\bf x}^\prime$ on the detector's trajectory which are connected by a light
ray reflected by the mirror. The essential fact is that the horizontal piece of the
curves is missing, which explains why the long-term oscillatory behaviour does not appear
in Fig.~4. The mechanism is that the additional logarithmic contribution besides the free
field term in the Wightman function (\ref{wiggen}) for large $s$ approximately equals minus
this term, which eliminates the horizontal piece and the tail of the curves. The rapid
decrease at $s\rightarrow \infty$ leads to a well-defined rate for $\tau\rightarrow \infty$
even without a decoupling factor $\eta=0$. The rates in Fig.~6 and 7 are obtained with this
value for $\eta$. Another important property is that $W(\tau, s)$ for $\tau\rightarrow \infty$
reproduces the quantity defined by the Wightman function for the uniformly accelerated
trajectories (\ref{d0})-(\ref{d2}), as can be also seen from the plot. This allows to
conclude that the KMS property reemerges in the limit of large acceleration times.

\begin{figure}[H]

\centering

\includegraphics[width=0.9\linewidth]{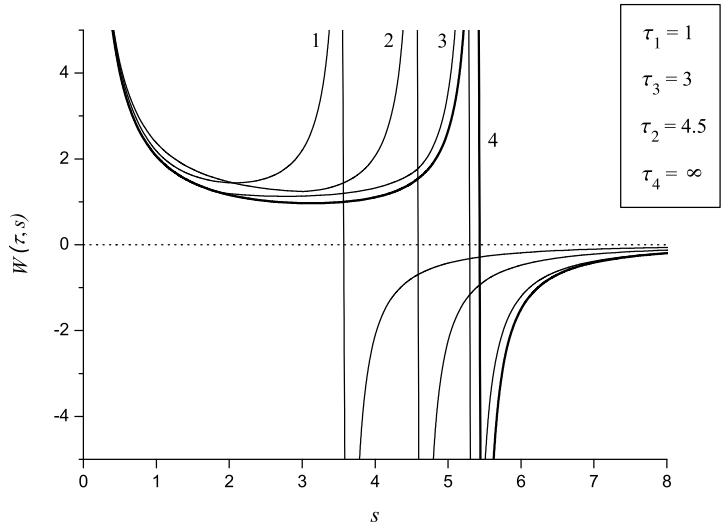}

\caption{The same as in Fig.~11 but for the detector in the presence of the mirror with
trajectories of the two objects the same as in Fig.~4. The curves for $\tau\rightarrow \infty$ indefinitely
approach curve $4$ (solid line), which corresponds to the Wightman function for infinite
acceleration times (\ref{d0})-(\ref{d2}).}

\end{figure}

\section*{Appendix C}

\setcounter{equation}{0}

\renewcommand{\theequation}{C.\arabic{equation}}

We arrive in this section to an approximate expression for (\ref{numsme}) that  is well suited
for numerical calculations. We choose the smearing functions to be unit-normalized rectangular
wave packets of the form
\bq
\qquad f_{\Omega}(\omega)=
\left
\{
\begin{array}{ll}
1/\sqrt{2\eta} \quad &\mbox{if}\quad
\omega \in [\Omega-\eta ,\, \Omega +\eta],
\\
0&\mbox{otherwise},
\end{array}
\right.
\label{smefun}
\eq
with $\Omega$ the mean frequency in the wave packet and $\Delta \omega=2\eta$
the frequency width. In these conditions the smeared beta coefficients are
\bq
\tilde \beta(\Omega, \omega^\prime)=
\int_{\Omega-\eta}^{\Omega+\eta}
d\omega \, \tilde \beta(\omega, \omega^\prime).
\label{smenew}
\eq
It is useful to parameterize the primed frequencies as
$\omega^\prime = a e^{a \sigma}$ with $\sigma\in(-\infty, \infty)$,
which brings (\ref{numsme}) to the form
\bq
n(\Omega_1, \Omega_2)= a^2
\int_{-\infty}^\infty d\sigma\, e^{a\sigma}
\tilde \beta (\Omega_1, \sigma)\, \tilde \beta(\Omega_2, \sigma)^*.
\label{intsig}
\eq
An obvious way to simplify (\ref{smenew}) is to assume very small frequency widths
$\eta \ll \Omega_1,\,\Omega_2$ and $a$. Starting with (\ref{newbet}) an evident
approximation then is (we replace in the front factor $\sqrt \omega\rightarrow \sqrt\Omega$)
\bq
e^{a \sigma/2}\tilde \beta(\Omega, \sigma)\simeq
-\frac{1}{2\pi}
\sqrt {\frac{\Omega}{a}}
\left \{b_1 (\Omega,\sigma)+b_{2} (\Omega,\sigma) \right\},
\quad
\label{sum2b}
\\
b_{1,2} (\Omega, \sigma)=
\frac{1}{\sqrt{2\eta}}
\int_{\Omega-\eta}^{\Omega+\eta}
d\omega \, b_{1,2}(\omega, \sigma).
\qquad\qquad\,\,
\label{betsme}
\eq
Similar approximations in (\ref{betsme}) must be operated with care due to the mixing
between $\omega$ and the unbounded variable $\sigma$, which can lead to non-negligible
variations of the integrands even for small variations of $\omega$. Let us introduce
\bq
\zeta=\sigma-\xi_0,
\quad
G(\zeta, \omega)=\int_0^\infty(e^{a \zeta}+it)^{i\omega/a} e^{-t}.
\label{gaminc}
\eq
Expressing (\ref{bb1}) and (\ref{bb2}) using (\ref{gaminc}) one finds
\bq
b_1 (\Omega,\sigma)&=&\frac{i}{\sqrt{2\eta}\,\Omega}
\int_{\Omega-\eta}^{\Omega+\eta} d\omega\,
e^{i\omega\sigma}
G^*(\sigma-\xi_0,\omega),
\label{c1}
\\
b_{2} (\Omega,\sigma) &=& -\frac{i}{\sqrt{2\eta}\,\Omega}\,
\int_{\Omega-\eta}^{\Omega+\eta} d\omega\,
e^{2i\omega \xi_0}\times e^{-i\omega\sigma}
G(\sigma-\xi_0,
\omega),
\label{c2}
\eq
where we eliminated the non-integral terms in (\ref{bb1}) and (\ref{bb2}), which cancel out
in (\ref{sum2b}). Note that we cannot make $\omega\rightarrow \Omega$ in the factors
$e^{\pm i\omega\sigma}$ due to the mixed dependence $\omega \sigma$. A closer look at
(\ref{gaminc}) shows that the same is valid in $G(\sigma-\xi_0, \omega)$. In order to
approximate these functions, the first step is to isolate the source of the rapid variations
with $\omega$. A clue is provided by the limits
\bq
\lim_{\zeta \rightarrow \infty} G(\zeta,\, \omega)
=e^{i\omega\zeta},
\quad
\lim_{\zeta \rightarrow -\infty} G(\zeta,\, \omega)
=e^{-\pi\omega/2a}\, \Gamma (i\omega/a+1).
\label{l2}
\eq
The expressions indicate that the rapid variations are contained only in the phase of
$G(\zeta, \omega)$, which is indeed confirmed by numerical calculations. With this
simplification and considering a sufficiently small $\eta$ we can approximate
\bq
G(\zeta,\, \omega)\simeq G(\zeta,\, \Omega)\,e^{i\Phi_{\Omega}(\zeta) (\omega-\Omega)},
\quad
\Phi_\Omega(\zeta)=\frac{\partial}{\partial \Omega}
\mbox{Arg}\, G(\zeta, \Omega).
\label{aprarg}
\eq
Note\, from (\ref{l2}) that $\lim_{\zeta \rightarrow -\infty} \Phi_\Omega(\zeta)\equiv \Phi(\Omega)$
remains finite. Introducing (\ref{aprarg}) in (\ref{c1}) and (\ref{c2}) the $\omega$-integrals
become trivial and one finds
\bq
b_1(\Omega, \sigma)&\simeq&
\sqrt{\frac{2}{\eta}}
\frac{i}{\Omega}\,
e^{i\Omega \sigma}\times
\frac{\sin[\eta(\sigma-\Phi_{\Omega}(\zeta)]}
{\sigma-\Phi_{\,\Omega}(\zeta)}\,
G^*(\zeta,\Omega),
\label{bG1}
\\
b_{2}(\Omega, \sigma)&\simeq&-
\sqrt{\frac{2}{\eta}}
\frac{i}{\Omega}\,e^{2i\Omega \xi_0-i\Omega\sigma}
\times
\frac{\sin [\eta (2\xi_0-\sigma+\Phi_{\,\Omega}(\zeta)]}
{2\xi_0-\sigma+\Phi_{\,\Omega}(\zeta)}\,
G(\zeta,\Omega).
\quad
\label{bG2}
\eq
It is now possible to explicitly write (\ref{intsig}). Using (\ref{sum2b}) the integral
can be organized as (the notation is obvious)
\bq
n = n_{11} +n_{22} +n_{12},
\label{n123}
\eq
where the three terms are completely defined by (\ref{bG1}) and (\ref{bG2}). We write
here only the first term:
\bq
 n_{11} (\Omega_1, \Omega_2)
\simeq
\frac{a}{2\pi^2 \eta\, \sqrt{\Omega_1\Omega_2}}
\int_{-\infty}^\infty d\sigma\,
G^*(\zeta, \Omega_1)\,G(\zeta, \Omega_2)
\qquad\quad\,\,\,
\nonumber
\\
\times
e^{i(\Omega_1-\Omega_2)\sigma}
\frac{\sin[\eta(\sigma-\Phi_{\Omega_1}(\zeta)]}
{\sigma-\Phi_{\Omega_1}(\zeta)}\,
\frac{\sin[\eta(\sigma-\Phi_{\Omega_2}(\zeta)]}
{\sigma-\Phi_{\Omega_2}(\zeta)}.\,\,
\label{n1ex}
\eq
The plots in Sec.~4.2.2 are based on the numerical evaluation of (\ref{n123}). It is essential to
mention that the three integrals in (\ref{n123}) separately diverge, which can be seen from the
fact that for $\sigma \rightarrow \infty$ the two terms (\ref{c1}) and (\ref{c2}) become independent
of $\sigma$, which follows from (\ref{l2}). However, these divergences are just an artefact of our
calculation. The problematic limit corresponds to $\omega^\prime\rightarrow \infty$ and going back to
(\ref{bb1}) and (\ref{bb2}) one can easily check that for $\omega^\prime$ large the beta coefficients
behave as $\sim 1/{\omega^\prime}^{\,3}$, which assures that (\ref{numsme}) remains finite. In the
numerical calculations one has to pay attention to this fact by evaluating all three integrals
with a common upper integration limit $\sigma_{max}\rightarrow \infty$.

We can now show that in the limit $\xi_0\rightarrow \infty$ we recover with (\ref{n123}) the
thermal expectation values defined by (\ref{smeter}). Although the $\sigma$-integrals
%with respect to $\sigma$
diverge,
the fact that the total integral converges still allows to let in the
integrands $\xi_0\rightarrow \infty$ keeping $\sigma$ fixed. Note that this implies
$\zeta \rightarrow -\infty$. It is then immediate from (\ref{bG2}) that
\bq
\lim_{\xi_0\rightarrow \infty} n_{22}(\Omega_1, \Omega_2)
=
\lim_{\xi_0\rightarrow \infty} n_{12}(\Omega_1, \Omega_2)=0.
\eq
It order to find the limit for $n_{11}$, we let $\zeta\rightarrow -\infty$ in (\ref{n1ex}) and
use the second limit in (\ref{l2}), which leads to
\bq
\lim_{\xi_0\rightarrow \infty} n (\Omega_1, \Omega_2)
=
\frac{a}{2\pi^2 \eta\,  \sqrt{\Omega_1\Omega_2}}\,
e^{-\pi(\Omega_1+\omega_2)/2a}\,
\Gamma (-i\Omega_1/a+1) \Gamma (i\Omega_2/a+1)\,
\nonumber
\\
\times
\int_{-\infty}^\infty d\sigma\, e^{i (\Omega_1-\Omega_2)\sigma}\,
\frac{\sin[\eta(\sigma-\Phi(\Omega_1)]}
{\sigma-\Phi(\Omega_1)}\, \frac{\sin[\eta(\sigma-\Phi(\Omega_2)]}
{\sigma-\Phi(\Omega_2)}.
\qquad\,\,\,\,
\label{n1li}
\eq
The trick is to recognize that the integral with respect to $\sigma$ is
\bq
\int_{-\infty}^\infty d\sigma\,(\dots)= \frac{1}{4}
\int_{-\infty}^\infty d\sigma
\int_{\Omega_1-\eta}^{\Omega_1+\eta}
d\omega_1
\int_{\Omega_2-\eta}^{\Omega_2+\eta}
d\omega_2\,
e^{i (\omega_1-\omega_2)\sigma}
\qquad\,
\\
\times
e^{-i \Phi(\Omega_1)(\omega_1-\Omega_1)
+i \Phi (\Omega_2)(\omega_2-\Omega_2)}
\simeq \frac{\pi}{2}\, \delta(\omega_1-\omega_2).
\label{dream}
\eq
The second relation follows from performing first the integral with respect to $\sigma$
and ignoring the second exponential factor, which is possible since the exponent is
$\sim \eta$. Using (\ref{dream}) one finally finds
\bq
\lim_{\xi_0\rightarrow \infty} n(\Omega_1, \Omega_2)
\simeq \frac{a}{4\pi \eta\,  \sqrt{\Omega_1\Omega_2}}\,
e^{-\pi(\Omega_1+\Omega_2)/2a}\,
\Gamma (-i\Omega_1/a+1) \Gamma (i\Omega_2/a+1)\,
\nonumber
\\
\times \int_{\Omega_1-\eta}^{\Omega_1+\eta}
d\omega_1
\int_{\Omega_2-\eta}^{\Omega_2+\eta}
d\omega_2\,
\delta(\omega_1-\omega_2).
\qquad\qquad\qquad\qquad\,
\label{endres}
\eq
It is easy to check that (\ref{endres}) approximates the thermal expectation values
$\langle {\bar a}^{R+}_{\,\Omega_1} \bar a^R_{\,\Omega_2}\rangle$ defined by (\ref{delbet})
smeared with the same functions (\ref{smefun}), assuming a sufficiently small $\eta$. The
exact thermal values differ from (\ref{endres}) only in that all factors above
fall under the integrals with the mean frequencies $\Omega_{1,2}$ replaced by
the integration variables $\omega_{1,2}$. The difference between the two quantities
can be seen in Fig.~8 and 9.
%We stress that although the function $\Phi(\Omega)$ plays
%no role in the limit $\xi_0\rightarrow \infty$, it is essential to include it in
%(\ref{bG1}) and (\ref{bG2}) for a valid approximation for $\xi_0$ finite.

\end{document}